\documentclass[aps,twocolumn,longbibliography]{revtex4-1}
\usepackage[]{graphicx}

\usepackage{color}

\usepackage{bm}

\usepackage{amssymb} 
\usepackage{amsmath}  % if you are using this package,it must be loaded before amsthm.sty
\usepackage{amsthm}
\usepackage{lgreek}

\begin{document}

\title{Dynamical systems with multiplicative noise: \\
Time-scale competition, delayed feedback and effective drifts}

\author{Giovanni Volpe}
\email{giovanni.volpe@fen.bilkent.edu.tr}
\affiliation{Soft Matter Lab, Department of Physics, and UNAM -- National Nanotechnology Research Center, Bilkent University, Ankara 06800, Turkey.}

\author{Jan Wehr}
\email{wehr@math.arizona.edu}
\affiliation{Department of Mathematics and Program in Applied Mathematics, University of Arizona, Tucson, Arizona 85721, USA.}

\date{\today}

\begin{abstract}
Noisy dynamical models are employed to describe a wide range of phenomena. Since exact modeling of these phenomena requires access to their microscopic dynamics, whose time scales are typically much shorter than the observable time scales, there is often need to resort to effective mathematical models such as stochastic differential equations (SDEs).  In particular, here we consider effective SDEs describing the behavior of systems in the limits when natural time scales became very small.  In the presence of {\it multiplicative noise} (i.e., noise whose intensity depends upon the system's state), an additional drift term, called {\it noise-induced drift}, appears. The nature of this noise-induced drift has been recently the subject of a growing number of theoretical and experimental studies. Here, we provide an extensive review of the state of the art in this field. After an introduction, we discuss a minimal model of how multiplicative noise affects the evolution of a system. Next, we consider several case studies with a focus on recent experiments:  Brownian motion of a microscopic particle in thermal equilibrium with a heat bath in the presence of a diffusion gradient, and the limiting behavior of a system driven by a colored noise modulated by a multiplicative feedback. This allows us to present the experimental results, as well as mathematical methods and numerical techniques that can be employed to study a wide range of systems. At the end we give an application-oriented overview of future projects involving noise-induced drifts, including both theory and experiment.
\end{abstract}

\maketitle

\section{Introduction}

\begin{figure}[b]
\includegraphics*[width=3.25in]{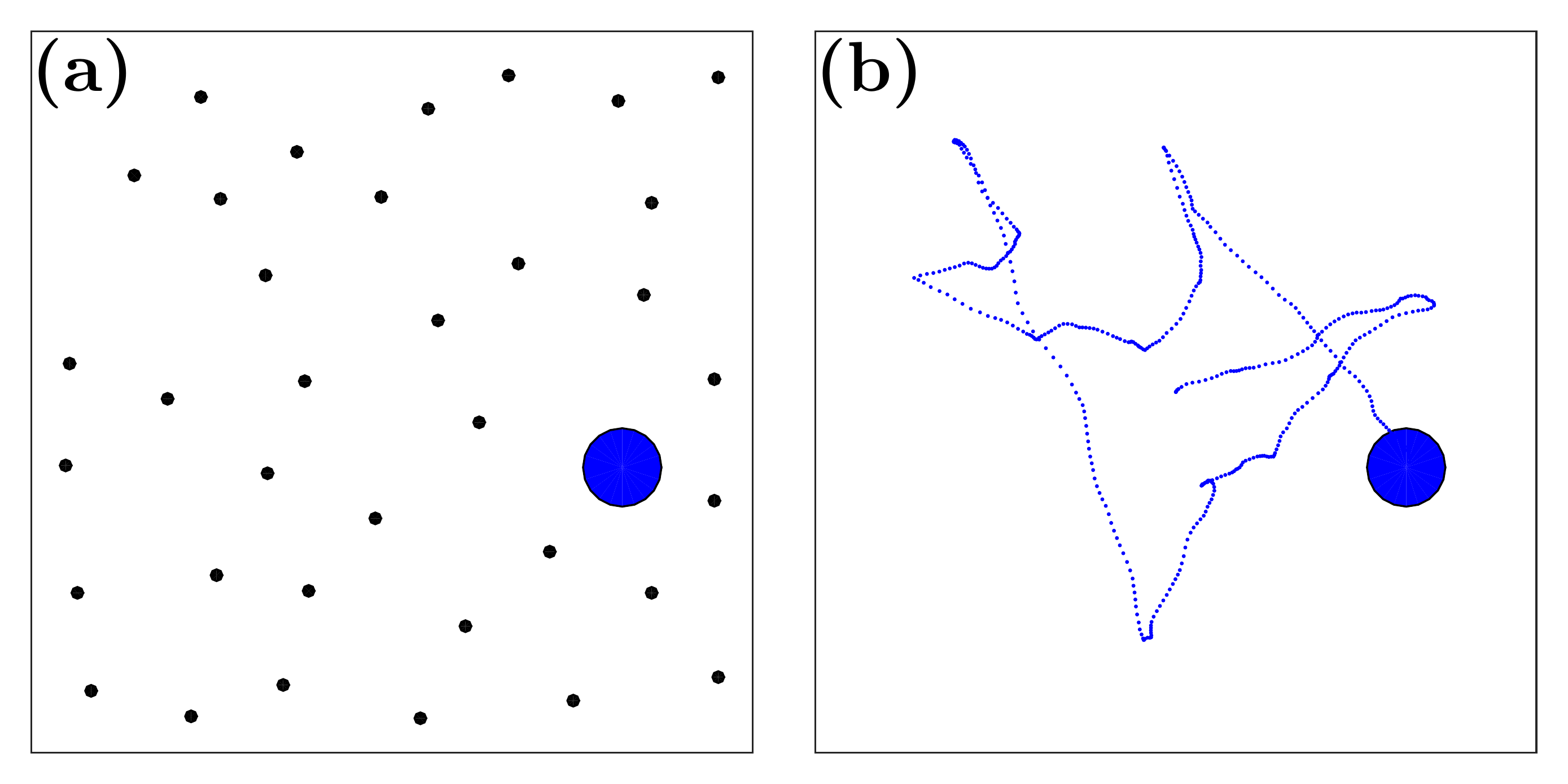}
\caption{(Color online) Stochastic motion from deterministic simulations. (a) A microscopic particle (large circle) immersed in a fluid continuously undergoes collisions with the fluid molecules (dots). (b) The resulting motion obtained from a molecular dynamics simulation (dotted line), despite being deterministic, appears to be random, especially if one has no access to the exact positions and velocities of the fluid molecules.}
\label{fig1}
\end{figure}

\begin{figure*}[t!]
\includegraphics*[width=\textwidth]{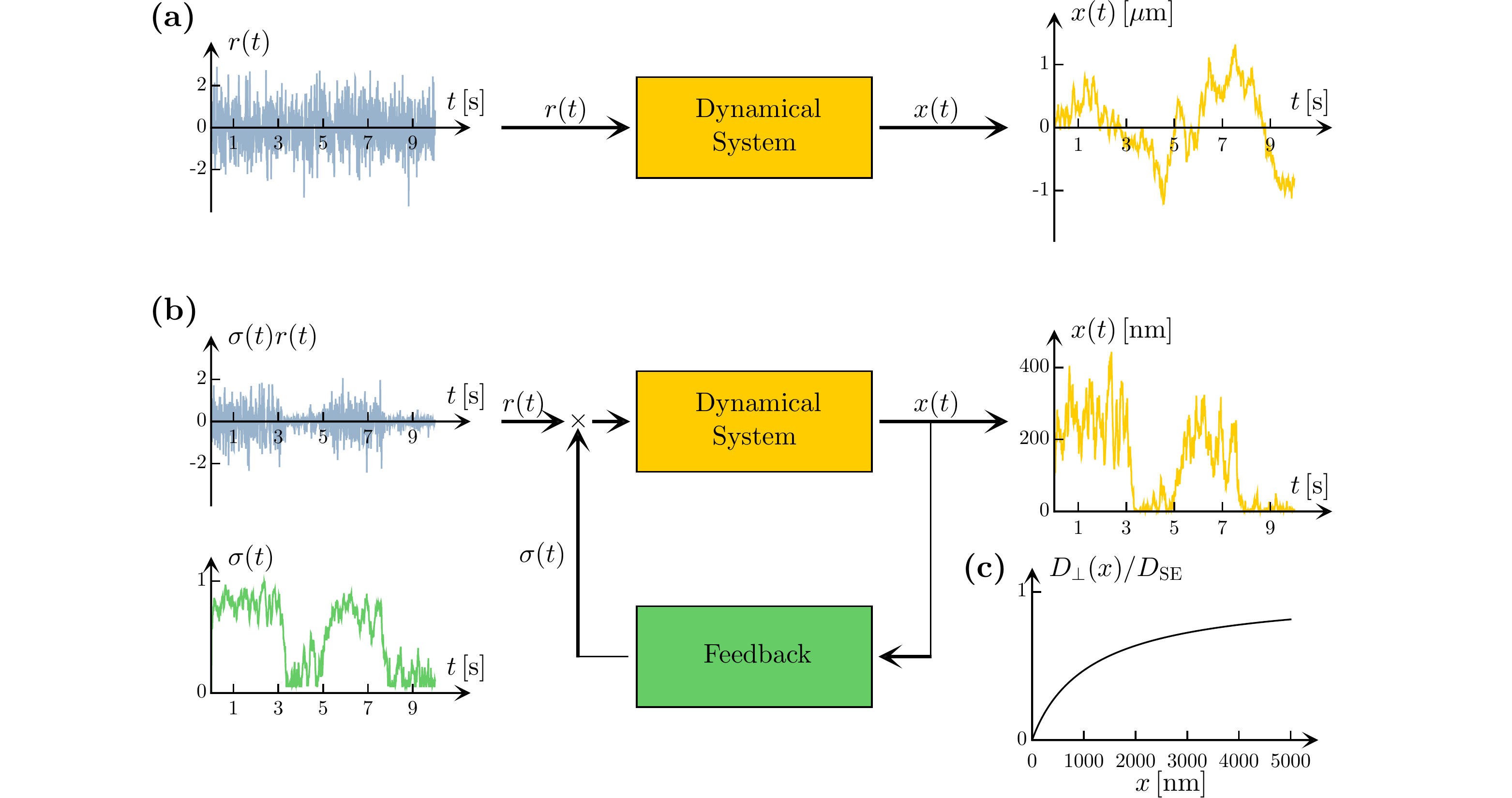}
\caption{(Color online) Stochastic dynamical system without and with feedback. (a) A schematic respresentation of a stochastic dynamical system: the system's status $x(t)$ evolves as the system is driven by a noisy input $r(t)$. (b) Same system with feedback $\sigma(x)$: $r(t)$ is now modulated by $\sigma(t)$, and $x(t)$ is clearly affected. The data correspond to the motion of a $1\,{\rm \mu m}$-radius Brownian particle in water at room temperature (a) in bulk and (b) close to a boundary; the curve in (c) shows the diffusion coefficient $D_{\perp}(x)$ of the particle in the direction perpendicular to the boundary (normalized to the bulk diffusion coefficient $D_{\rm SE} = D_{\perp}(\infty)$) as a function of its distance from the boundary $x$.}
\label{fig2}
\end{figure*}

Dynamical systems are widely employed to describe and predict the behavior of complex phenomena \cite{strogatz2014nonlinear}. At any given time $t$, a dynamical system is characterized by a state $x_t$, which evolves according to a deterministic rule. A complete deterministic description requires access to the system's microscopic dynamics. A classical example is the Brownian motion of a particle in a fluid \cite{nelson1967dynamical}. The motion of the particle and fluid molecules is deterministic, as it can be described by a set of Newton's equations: knowing the initial positions and velocities of all particles (i.e., the Brownian particle and the fluid molecules), it is in principle possible to determine their motion over time, as is done in the molecular dynamics simulation shown in Fig.~\ref{fig1}a. Nevertheless, the resulting motion of the microscopic particle (shown in Fig.~\ref{fig1}b) appears to be random, especially if one has no access to the exact positions and velocities of the fluid molecules. In fact, it is often an impossible task to construct a model for a dynamical system that accounts for its microscopic dynamics. For example, even though in principle it would be possible to construct a model of Brownian motion writing down Newton's equation of motion for the particle as well as for each fluid molecule, this is a practically unfeasible task due to the huge number of molecules in any realistic situation --- of the order of the Avogadro number $6.02 \cdot 10^{23}$. 

It is often convenient to reduce the effective number of degrees of freedom in order to obtain more tractable models. This can be achieved by introducing  some randomness. For example, the Brownian motion of a particle can be modelled by the stochastic differential equation (SDE)
\begin{equation}\label{eq:freebrownian}
dx_t = \sigma \, dW_t \; ,
\end{equation}
where $W_t$ is a Wiener process, i.e., a stochastic process with continuous paths, whose increments $W_t - W_s$ are independent and normally distributed with mean zero and variance $t-s$ \cite{karatzas2012brownian}, representing the stochastic driving, $\sigma = \sqrt{2D_{\rm SE}}$ is a constant, and $D_{\rm SE}$ is the Stokes-Einstein diffusion constant \cite{nelson1967dynamical}. The resulting Brownian motion is shown in Fig.~\ref{fig2}a for the case of a $1\,{\rm \mu m}$-radius Brownian particle in water at room temperature in bulk, i.e., far away from any boundaries. SDE~(\ref{eq:freebrownian}) is arguably the simplest way to describe the properties of a free diffusion, as it only involves explicitly one degree of freedom. The term $\sigma \, dW_t$ is thus a mathematical model for the noise, which permits one to reduce the number of degrees of freedom and to implicitly account for the microscopic dynamics of the system (in this case, the motion of the fluid molecules). We stress that SDE~(\ref{eq:freebrownian}) with an initial condition $x_0$ has a unique solution, $x_t = x_0 + \sigma W_t$, for any given realization of $W_t$. Similar models have been employed to describe a wide range of phenomena, from thermal fluctuations in electronic circuits and the evolution of stock prices, to heterogeneous response of biological systems to stochasticity in gene expression \cite{kloeden1992numerical,oksendal2013stochastic}.

Often, the system's state influences the intensity of the driving noise, as it is fed back on the input noise and modulates it. If such feedback loop is multiplicative, i.e., the intensity of the input noise gets \emph{multiplied} by a function of the system's state, as shown in Fig.~\ref{fig2}b, one says that the system is driven by a \emph{multiplicative noise}. For example, the Brownian fluctuations of a microscopic particle near a wall are reduced by hydrodynamic interactions \cite{brenner1961slow}, as shown in Fig.~\ref{fig2}c. The corresponding SDE (in the absence of other forces, see Eq.~(\ref{eq:diffusiongradient2}) for a more complete model) is
\begin{equation}\label{eq:diffusiongradient}
dx_t = \sigma(x_t) \, dW_t \; ,
\end{equation}
where $\sigma(x_t) = \sqrt{2D_{\perp}(x_t)}$ and $D_{\perp}(x)$ is the particle's diffusion coefficient in the direction normal to the wall, which depends on the particle-wall distance $x$. Similar models are employed to describe, e.g., the change of the step size of a random walk due to inhomogeneity of the medium \cite{lanccon2001drift}, the alteration of the volatility of a stock price depending on its actual value \cite{hamao1990correlations}, and the regulation of the stochastic expression of a gene by the concentration of its products \cite{kaern2005stochasticity}.

Unlike SDE~(\ref{eq:freebrownian}), the integration of SDE~(\ref{eq:diffusiongradient}) has to be performed carefully, because a realization of the Wiener process $W_t$ has infinite variation on any interval (in fact, the derivative $dW_t \over dt$ does not exist anywhere) \cite{karatzas2012brownian}. The stochastic integral $\int_{0}^{T} f(x_t) \circ_\alpha dW_t \equiv  \lim_{N \to \infty} \sum_{n=0}^{N - 1} f(x_{t_n})\Delta W_{t_n}$, where $t_n = \frac{n+\alpha}{N} T$ and $\alpha$ is a real number (typically, $\alpha = 0$, $0.5$ or $1$), may have different values for different choices of $\alpha$ \cite{sussmann1978gap,karatzas2012brownian}. Therefore, a complete model is defined by an SDE and the integration convention, which must be determined on the basis of the available experimental data or derived from another unambiguous model \cite{van1981ito}. If desired, one can change the convention to $\alpha'$, but only by adding an appropriate \emph{noise-induced drift} term at the same time; as we will see in Section~\ref{sec:simple}, this noise-induced drift term is in general proportional to $\sigma(x_t) \frac{d}{dx} \sigma(x_t)$ \cite{karatzas2012brownian}. Thus, a more precise way of writing SDE~(\ref{eq:diffusiongradient}) is
\begin{equation}\label{eq:ab}
dx_t = (\alpha-\alpha') \sigma(x_t) \frac{d \sigma(x_t)}{dx} \, dt + \sigma(x_t) \circ_{\alpha'} dW_t \; ,
\end{equation}
where the integration convention indicated by $\alpha'$ and the noise-induced drift, i.e., $(\alpha-\alpha') \sigma(x_t) \frac{d \sigma(x_t)}{dy}$, are explicitly shown. This shows that the equations $dx_t = \sigma(x_t) \circ_{\alpha} dW_t$ are not equivalent for different $\alpha$ and it is thus clear that from the modeling perspective the choice of the appropriate SDE-convention pair is of critical importance, especially when the model is employed to predict the system's behavior under new conditions.

Finally, let us note that until now we have only considered equations without a deterministic drift. If a deterministic drift $g(x_t)$ is present, SDE~(\ref{eq:diffusiongradient}) becomes
\begin{equation}\label{eq:diffusiongradient2}
dx_t = g(x_t) \, dt + \sigma(x_t) \, dW_t \; .
\end{equation}
However, the presence of $g(x_t) \, dt$ does not lead to any ambiguities, since this term can be integrated in a standard way.

In Section~\ref{sec:simple}, we introduce the fundamental concepts and ideas in a simple and intuitive way, making use of a minimal discrete-time model. In Section~\ref{sec:cases}, we describe in detail some case studies focusing mainly on recent experiments; this allows us to present not only the experimental findings, but also some mathematical methods and numerical techniques that can be employed to study a wide range of systems. Finally, in Section~\ref{sec:outlook}, we give an overview of various other situations where noise-induced drifts in the limiting SDEs, describing a system driven by multiplicative noise, become relevant. We argue that the possibility of such noise-induced drifts and of their dramatic consequences should be recognized and accounted for in many cases where SDEs with multiplicative noise are routinely employed to predict the behavior and evolution of complex physical, chemical, biological and economic phenomena. We conclude with some perspectives for future developments of this field.

\section{A minimal discrete-time model}\label{sec:simple}

In this section, we introduce a minimal (discrete-time) model to demonstrate how multiplicative noise affects the evolution of a system. We will, in particular, see how the presence of a multiplicative noise can generate a noise-induced drift and alter the long-term probability distribution of the system's state. 

\begin{figure*}[t!]
\includegraphics*[width=6in]{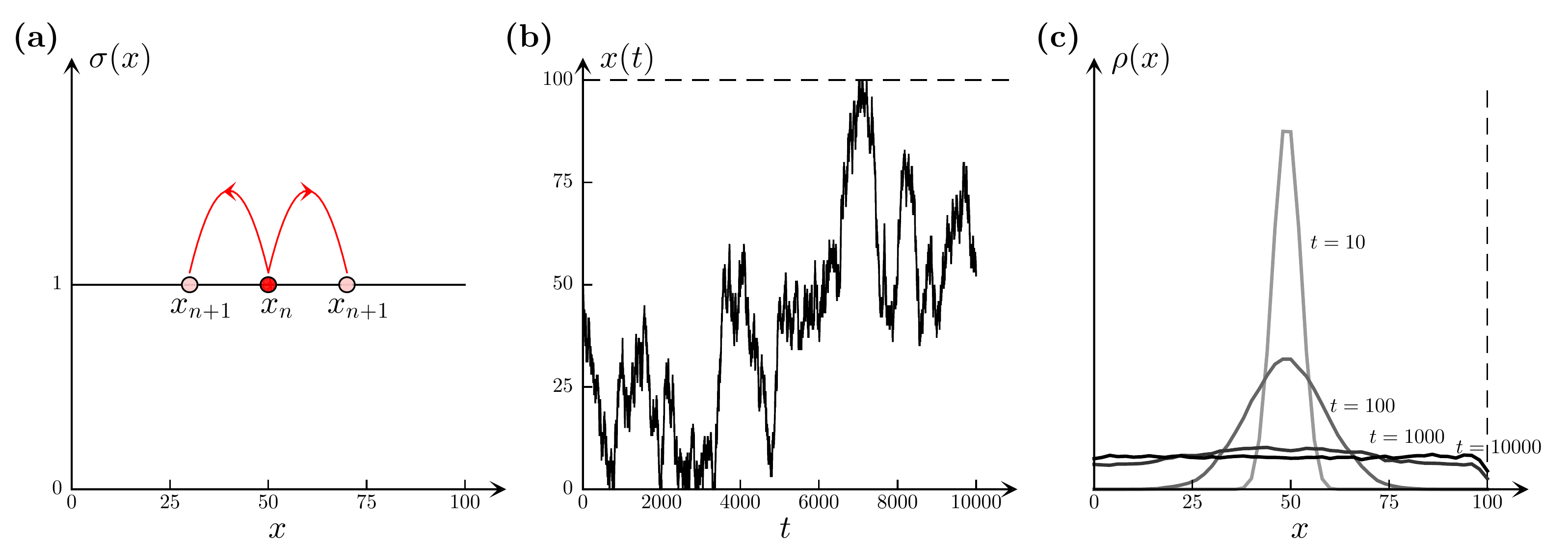}
\caption{(Color online) Evolution of the dynamical system described by SDE~(\ref{eq:FD1}). (a) The intensity of the noise $\sigma$ does not depend on the system's state $x$; therefore at each time step the state increases or decreases by a fixed amount $\sigma \sqrt{\Delta t}$ (not to scale) with equal probability (0.5). (b) Example of a trajectory of the system in state space ($\sigma=1$). (c) Probability density distributions at selected times (calculated from $10\,000$ simulated trajectories). Reflecting boundary conditions are imposed at $x=0$ and $x=100$. Note that the steady-state probability distribution is uniform, as can be expected from the absence of deterministic forces acting on the system.}
\label{fig3}
\end{figure*}

We start by considering the system without multiplicative noise described by SDE~(\ref{eq:freebrownian}). The continuous-time solution $x(t)$ of SDE~(\ref{eq:freebrownian}) can be approximated by a discrete-time sequence $x_n$, which is the solution of the corresponding finite-difference equation (FDE) evaluated at regular time steps $t_n = n \Delta t$; for $\Delta t$ sufficiently small, $x_n \approx x(t_n)$. The finite-difference (FD) terms corresponding to $dx_t$ are $x_{n+1} - x_n$, while those corresponding to $dW_t$ are given by a sequence of independent random numbers with zero mean and variance $\Delta t$ \footnote{This follows from the properties of a Wiener process $W_t$: $\left< \int_t^{t+\Delta t} dW_t \right> = \left< W_{\Delta t} \right> = 0$ and $\left< \left( \int_t^{t+\Delta t} dW_t \right)^2  \right> =  \left< W_{\Delta t}^2 \right> = \Delta t$. A more detailed discussion can be found in \citet{volpe2013simulation}.}, such as a sequence of indepenendent random numbers with values $\pm \sqrt{\Delta t}$. We thus obtain the discrete-time random walker FDE:
\begin{equation}\label{eq:FD1}
x_{n+1} = x_n \pm \sigma \, \sqrt{\Delta t} \; ,
\end{equation}
where the symbol ``$\pm$'' signifies that at each step the sign is chosen randomly. As shown in Fig.~\ref{fig3}a, at each time step the value of the system's state either increases or decreases with the same probability ($0.5$) and amplitude ($\sigma \sqrt{\Delta t}$). In Fig.~\ref{fig3}b, we show a simulated trajectory for the evolution of such system starting at $x_0 = 50$. Since the probability and amplitude of the step are equal in both directions (i.e., ``$+$" and ``$-$"), the system's state evolves in a symmetric way. In the simulations presented in Fig.~\ref{fig3}, in order to be able to obtain a steady-state probability distribution for the system's state, we have restricted the system's space to the interval between $0$ and $100$, introducing reflecting boundary conditions at $x=0$ and $x=100$ \footnote{In absence of boundary conditions, the state space would be unbounded and a steady state probability distribution would not exist.}. As shown in Fig.~\ref{fig3}c, we obtain a steady-state probability distribution that is uniform, as can be expected due to the absence of deterministic forces acting on the system \cite{lanccon2001drift}.

\begin{figure*}[t!]
\includegraphics*[width=6in]{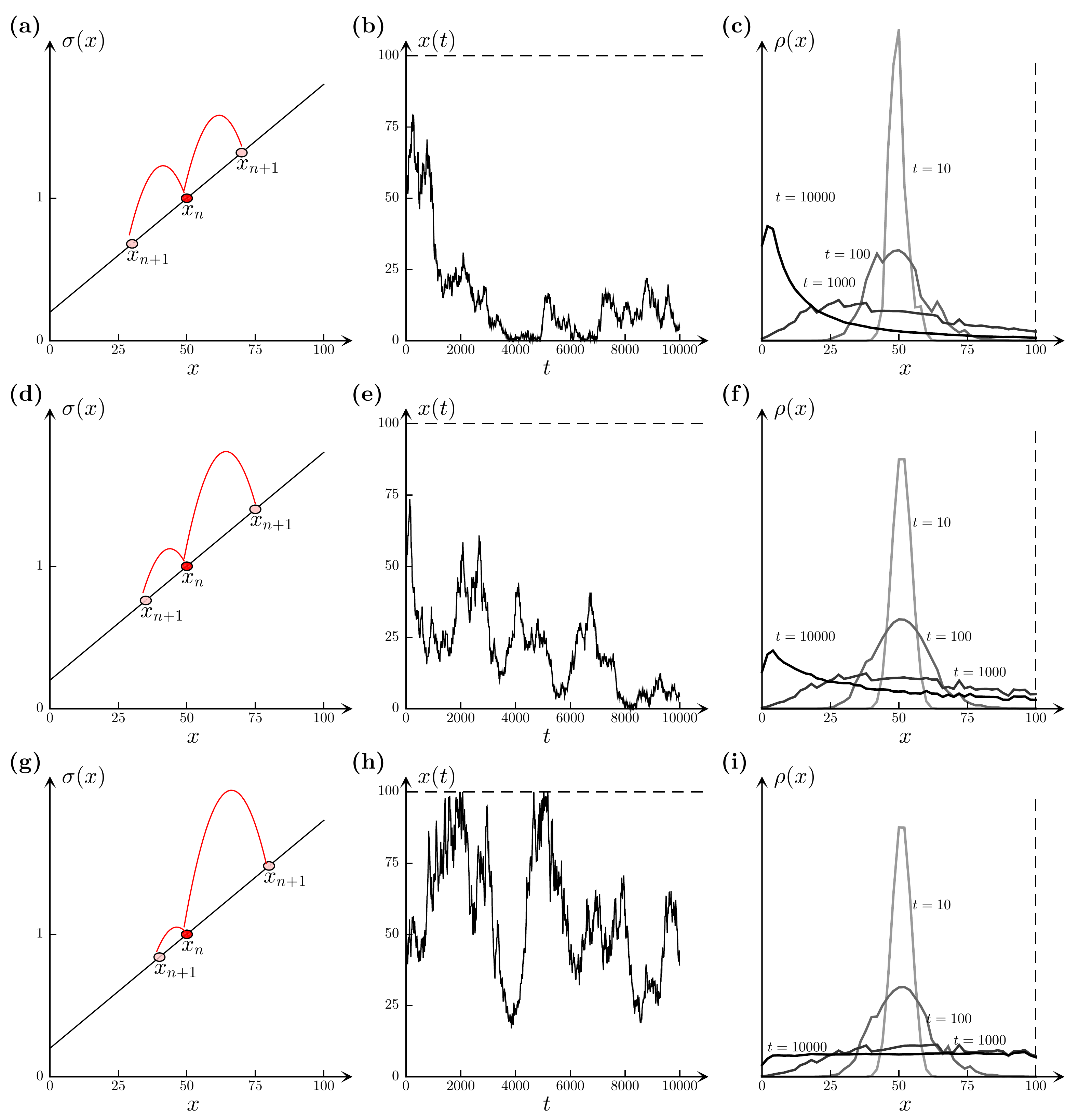}
\caption{(Color online)  Evolution of the dynamical system with multiplicative noise described by SDE~(\ref{eq:a}) for various values of $\alpha$. (a) For $\alpha=0$ (Eq.~(\ref{eq:0})), the amplitude of each random step is a function of the initial state and is therefore symmetrically distributed; (b) example of a trajectory in state space; (c) probability density distributions at selected times. The corresponding results for $\alpha=0.5$ (Eq.~(\ref{eq:0.5})) and $\alpha=1$ (Eq.~(\ref{eq:1})) are shown in (d-f) and (g-i), respectively. In all cases, reflecting boundary conditions are imposed at $x=0$ and $x=100$. Note that only in the $\alpha = 1$ case the steady-state probability distribution is uniform, while in the other two cases it is peaked in the low-noise (small $\sigma(x)$) region. The steady-state probability distributions are calculated from $100\,000$ simulated trajectories.}
\label{fig4}
\end{figure*}

We will now consider the system with multiplicative noise described by SDE~(\ref{eq:diffusiongradient}). Again, we can approximate the continuous-time solution by a discrete-time sequence of states, which solves the corresponding FDE. Now, however, we have to decide: where should the value of $\sigma(x)$ be evaluated at consecutive time steps? At the starting state $x_n$? At the final state $x_{n+1}$? At the (midpoint) intermediate state ${1\over2}(x_n + x_{n+1})$? At some other state?

Let us first consider the case when the value of $\sigma(x)$ is evaluated at $x_n$ (Figs.~\ref{fig4}a-c); explicitly:
\begin{equation}\label{eq:0}
x_{n+1} = x_n \pm \sigma(x_n) \, \sqrt{\Delta t} \; .
\end{equation}
This is particularly convenient from a computational point of view because the value of $x_n$ is already available when the FDE is solved iteratively (see also Appendix~\ref{app:sim}). As shown in Fig.~\ref{fig4}a, the value of $x$ either increases or decreases by the same amount, equal to $\sigma(x_n)\sqrt{\Delta t}$ ($\sigma(x)$ is plotted by the black solid line). A numerical solution is shown in Fig.~\ref{fig4}b and the evolution of the probability density of the system's state is shown in Fig.~\ref{fig4}c. At the beginning, the state is $x_0 = 50$ and evolves in a symmetric way, but, as time passes, the system reaches an asymmetric steady-state probability distribution and is more likely found in low-noise states, i.e, states for which $\sigma(x)$ is smaller.

We can also consider the case when the value of $\sigma(x)$ is evaluated at the midpoint state ${1\over2}(x_n + x_{n+1})$. A heuristic argument for applying this convention to real systems is that the value of $\sigma(x)$ should be averaged over the change of the system's state. In this case, the corresponding FDE is
\begin{equation}\label{eq:0.5}
x_{n+1} = x_n \pm \sigma\left({\textstyle {1\over2}}(x_n + x_{n+1})\right) \, \sqrt{\Delta t} \; .
\end{equation}
We will explain how to approximately solve this equation below (see Eq.~(\ref{eq:a2}) and Appendix~\ref{app:sim}). Fig.~\ref{fig4}d shows that the change of the system's state now becomes asymmetric because it is larger (smaller) when moving toward increasing (decreasing) $\sigma$. A simulated trajectory is shown in Fig.~\ref{fig4}e and the evolution of the probability density of the system's state is shown in Fig.~\ref{fig4}f. At the beginning the probability density drifts towards higher-noise states.
% ; this is a first glimpse of the presence of a noise-induced drift. 
However, at long times, the system is still more likely to be found in low-noise states.

Furthermore, for reasons that will become clear later (Section~\ref{sec:case:bm}), it can be also useful to evaluate $\sigma$ at other states and, in particular, at the final state $x_{n+1}$. In this case, the corresponding FDE is given by
\begin{equation}\label{eq:1}
x_{n+1} = x_n \pm \sigma(x_{n+1}) \, \sqrt{\Delta t} \; .
\end{equation}
The change in the system's state becomes even more asymmetric than in the previous case (Fig.~\ref{fig4}g) and an even larger noise-induced drift can be seen at short times (Figs.~\ref{fig4}h and \ref{fig4}i). Interestingly, the steady-state probability density distribution appears to be uniform, i.e., independent of the value of $\sigma(x)$. We can conclude that in this case the noise-induced drift is sufficient to compensate for the tendency of the system to linger in low-noise states.

In order to understand the origin of the noise-induced drift and how it is related to the way the noise term is evaluated, we study the following FDE:
\begin{equation}\label{eq:a}
x_{n+1} = x_n \pm \sigma(x_n + \alpha \Delta x) \, \sqrt{\Delta t} \; ,
\end{equation}
where $\Delta x = x_{n+1} - x_n$. We expand the factor $\sigma(x_n + \alpha \Delta x)$ as
$$
\sigma(x_n + \alpha \Delta x) \approx \sigma(x_n) + \alpha \frac{d\sigma(x_n)}{dx} \Delta x \; .
$$
Substituting the first-order expansion of $\Delta x \approx \pm \sigma(x_n) \sqrt{\Delta t}$, we obtain
$$
\sigma(x_n + \alpha \Delta x) \approx \sigma(x_n) \pm \alpha \, \sigma(x_n) \frac{d\sigma(x_n)}{dx} \sqrt{\Delta t} \; 
$$
and we can therefore re-write Eq.~(\ref{eq:a}) as
\begin{equation}\label{eq:a2}
x_{n+1} = x_n + \underbrace{\alpha \, \sigma(x_n) \frac{d\sigma(x_n)}{dx} \Delta t}_{\mbox{noise-induced drift}} \pm \sigma(x_n) \, \sqrt{\Delta t} \; .
\end{equation}
Therefore, various values of $\alpha$ lead to different noise-induced drifts and, consequently, to different steady-state probability distributions, as we have seen in Fig.~\ref{fig4} for the cases $\alpha = 0$, $0.5$ and $1$. Importantly, we note that the presence of the noise-iduced drift does not depend on the value of $\Delta t$, i.e., it is present in the limiting SDE as we will see in the case studies presented in the Section~\ref{sec:cases}. 

The parameter $\alpha$ determines how the stochastic integration is performed. Common choices are: the {\it It\^{o} integral} with $\alpha=0$ corresponding to the use of the initial value (Eq.~(\ref{eq:0})) \cite{ito1944109}; the {\it Stratonovich integral} with $\alpha=0.5$ corresponding to the use of the midpoint value (Eq.~(\ref{eq:0.5})) \cite{stratonovich1966new}; and the {\it anti-It\^{o}} or {\it isothermal integral} with $\alpha=1$ corresponding to the use of the final value (Eq.~(\ref{eq:1})) \cite{hanggi1982nonlinear,klimontovich1990ito}. In particular, $\alpha = 0$ models are typically employed in economics \cite{oksendal2013stochastic} and biology \cite{turelli1977random}, because of their property of ``not looking into the future,'' referring to the fact that, when the integral is approximated by a sum, the first point of each interval is used (see also Appendix~\ref{app:sim}). $\alpha = 0.5$ naturally emerges in physical systems with noise correlation time $\tau > 0$, e.g., the SDEs describing electrical circuits driven by a multiplicative noise \cite{smythe1983observation};  this is explained mathematically by the Wong-Zakai theorem, which states that, if in SDE~(\ref{eq:diffusiongradient}) the Wiener process is approximated by a sequence of smooth processes with symmetric covariance and $\tau$ going to $0$, the resulting limiting SDE should be interpreted according to Stratonovich calculus \cite{wong1969riemann,wong1969riemann}. Finally, $\alpha = 1$ naturally emerges in physical systems in equilibrium with a heat bath \cite{ermak1978brownian,lanccon2001drift,lau2007state,volpe2010influence}. In some dynamical systems, e.g., circuits with time delay and colored noise (see Section~\ref{sec:case:color}), $\alpha$ can actually vary under changing operational conditions \cite{pesce2013stratonovich}.

SDE~(\ref{eq:ab}) explicitly states the integration convection $\alpha'$ and the noise-induced drift $(\alpha - \alpha')$. If desired, one can change the convention ($\alpha'$), but this entails a corresponding change in the drift term ($\alpha - \alpha'$). For example, 
\begin{equation}
dx_t = \underbrace{\sigma(x_t) \frac{d \sigma(x_t)}{dy}}_{\alpha-\alpha' = 1} + \underbrace{\sigma(x_t) dW_t}_{\alpha' = 0} \; ,
\end{equation}
is equivalent to 
\begin{equation}
dx_t = \underbrace{0.5 \, \sigma(x_t) \frac{d \sigma(x_t)}{dy}}_{\alpha-\alpha' = 0.5} + \underbrace{\sigma(x_t) \circ dW_t}_{\alpha' = 0.5} \; ,
\end{equation}
and to
\begin{equation}
dx_t = \underbrace{\sigma(x_t) \circ_1 dW_t}_{\alpha-\alpha' = 0,\; \alpha' = 1} \; ,
\end{equation}
where we are using the common notations $\sigma(x_t) \, dW_t \equiv \sigma(x_t) \, \circ_0 \, dW_t$ and $\sigma(x_t) \, \circ \, dW_t \equiv \sigma(x_t) \, \circ_{0.5} \, dW_t$. In this review, unless otherwise stated, we will use the It\^o convention ($\alpha = 0$) throughout and explicitly indicate the noise-induced drifts to avoid misunderstandings associated with changing formalism.

Before moving to the case studies in next section, we want to make an important remark. In this section we have considered only first-order SDEs, where the presence of noise-induced drift is related to the choice of a stochastic integration convention. In the case studies in Section~\ref{sec:cases}, we will typically start from a microscopic model of a system and eliminate some of its complexity to obtain an effective first-order SDE. The noise-induced drift present in the effective first-order SDE will, thus, be the result of this simplification process. For clarity, we will always write the effective SDEs using the It\^o formalism, where the noise-induced drift is explicitly stated. For example, in Sections~\ref{sec:case:bm} and \ref{sec:case:fdr}, our starting point is a second-order equation, which we want to simplify further taking a parameter (e.g., mass of a particle) to zero.  The resulting first-order It\^o equation contains a drift term that combines the damping and the noise coefficients of the original equation.  We emphasize that the source of this (physically measurable) additional drift is that we are taking a singular limit of a second-order equation {\it in the presence of state-dependent noise} and we thus call it again a {it noise-induced drift}.  Its explicit form is now much harder to derive than in the case of the minimal model of Section~\ref{sec:simple}.  In the case discussed in Section~\ref{sec:case:bm}, it is possible (but not necessary) to interpret it in terms of a stochastic integration convention choice ($\alpha = 1$) \cite{mannella2011comment,volpe2011reply}, as explained in Section~\ref{sec:simple}, but no such interpretation is possible in the generality of Section~\ref{sec:cases}.   

\section{Case studies}\label{sec:cases}

In Section~\ref{sec:simple} we have seen how the presence of multiplicative noise induces a drift in a simple discrete-time model of a random walker. In the present section we consider in detail several examples of realistic models with a particular emphasis on those systems that have been subject of experiments. Section~\ref{sec:case:bm} considers Brownian motion of a microscopic particle in thermal equilibrium with a heat bath, i.e., for which the fluctuation-dissipation relations holds, in the presence of a diffusion gradient. Section~\ref{sec:case:fdr} relaxes the condition that the system should be in equilibrium with a heat bath and thus considers systems for which a generalized fluctuation-dissipation relation holds. Section~\ref{sec:case:color} considers the limiting behavior of a system driven by a colored noise modulated by a multiplicative delayed feedback. 
%Finally, in Section~\ref{sec:case:delay} we demonstrate how sensorial delay can alter the behavior of an autonomous agent in the presence of noise and how this effect can be used to control complex collective behaviors. 
In all cases we will present not only experimental findings, but also mathematical methods and/or numerical techniques that can be employed to study a wide range of systems.
 
 \subsection{Brownian motion in a diffusion gradient}\label{sec:case:bm}

Diffusion gradients emerge naturally when a Brownian particle is in a complex or crowded environment. For example, diffusion gets hindered when a particle is close to a wall due to hydrodynamic interactions: as shown in Fig.~\ref{fig2}c, the diffusion coefficient increases with the particle-wall distance approaching its bulk value at a distance of several particle radii away from the wall \cite{brenner1961slow}. The presence of a diffusion gradient introduces a multiplicative noise and thus leads to a noise-induced drift, often referred to in this context as a ``spurious drift". The need to account for such spurious drifts was realized already several decades ago in the context of numerical simulations \cite{ermak1978brownian,ryter1981brownian,hanggi1982nonlinear,sancho1982adiabatic}, but only very recently did it become possible to observe them experimentally \cite{lanccon2001drift,lanccon2002brownian,volpe2010influence,brettschneider2011force}.

In order to understand how spurious drifts emerge in the presence of diffusion gradients, we will consider a Brownian particle with mass $m$ moving in one dimension in a fluid at temperature $T$. Importantly, we assume that the particle is in thermal equilibrium with the heat bath provided by the fluid. The Newton's equation of motion is
\begin{equation}\label{eq:mass}
m\ddot{x}_t = F(x_t) - \gamma(x_t)\dot{x}_t + \gamma(x_t)\sqrt{2D(x_t)}\eta_t \; ,
\end{equation}
where $F(x)$ denotes the sum of the external forces acting on the particle, $\gamma(x)$ is the position-dependent friction coefficient, $D(x)$ is the position-dependent diffusion coefficient, and $\eta_t$ is a unit white noise. Since we assume that the system is in thermal equilibrium, the intensity of the fluctuations $D(x)$ and the rate of energy dissipation $\gamma(x)$ satisfy the fluctuation-dissipation relation \cite{zwanzig2001nonequilibrium}
\begin{equation}\label{eq:einstein}
D(x) = \frac{k_{\rm B}T}{\gamma(x)} \; ,
\end{equation}
where $k_{\rm B} T$ is the thermal energy and $k_{\rm B}$ is the Boltzmann constant.

SDE~(\ref{eq:mass}) is equivalent to the system
\begin{equation}\label{eq:system}
\left\{\begin{array}{ccl}
\displaystyle dx_t 
& = & 
\displaystyle v_t\,dt \; , \\[6pt]
\displaystyle dv_t 
& = & 
\displaystyle {F(x_t) \over m}\,dt - {k_{\rm B}T \over mD(x_t)}v_t\,dt + {k_{\rm B}T\sqrt{2} \over m\sqrt{D(x_t)}}\,dW_t \; ,
\end{array}\right.
\end{equation}
where $W_t = \int_0^t\eta_s\,ds$ is the time integral of the white noise \footnote{Note that in SDEs~(\ref{eq:system}) there is no multiplicative noise because the noise term for $v_t$ is multiplied by a function of $x_t$ (not of $v_t$).}.
The evolution of a probability density $\rho(x,v)$ under the stochastic dynamics defined by this system is described by the Fokker-Planck (in mathematics literature:  forward Kolmogorov) equation
\begin{equation}
\rho_t = {(k_{\rm B}T)^2 \over m^2D(x_t)}\rho_{vv} - v\rho_x - {F(x) \over m}\rho_v + {k_{\rm B}T \over mD(x)}(\rho v)_v \; ,
\end{equation}
where the subscripts denote partial derivatives. To find the steady-state probability density, we need to solve this equation with the left-hand side equal to zero.  A direct calculation shows that it is satisfied by the Boltzmann-Gibbs probability distribution, i.e.,
\begin{equation}\label{eq:BG}
\rho(x,v) = Z^{-1}\exp\left[-{U(x) \over k_{\rm B}T} - {mv^2 \over 2k_{\rm B}T}\right] \; ,
\end{equation}
where $U(x)$ is the potential of the (external) forces, i.e., $F(x) = -{dU(x) \over dx} $, and we are assuming that the density is normalizable with $Z$ denoting the normalizing factor. Furthermore, the Maxwellian velocity distribution ($\propto \exp\left[- {mv^2 \over 2k_{\rm B}T} \right]$) implies energy equipartition, so that the equilibrium kinetic energy is on average equal to the thermal energy, i.e.,
\begin{equation}\label{eq:maxwellian}
\left< mv_t^2 \right > = k_{\rm B}T \; .
\end{equation}

In SDE~(\ref{eq:mass}), inertial effects decay on a very short time scale, i.e., the \emph{momentum relaxation time} $\tau_{\rm m} = m/\gamma$, which is typically of the order of a fraction of a microsecond. For example, for a silica microsphere with radius $R = 1\,{\rm \mu m}$ ($m = 11\,{\rm pg}$) in water at room temperature ($T = 300\,{\rm K}$), $\tau_{\rm m} = 0.6\,{\rm \mu s}$. This time is several orders of magnitude shorter than the time scales of typical experiments, which are of the order of milliseconds or longer \footnote{In fact, recent experiments have been able to resolve the inertial regime of Brownian particles immersed both in a gas and in a liquid. For a recent review see \citet{li2013brownian}.} \footnote{In a liquid environment, furthermore, also the hydrodynamic memory of the fluid, i.e., the mass of the fluid displaced together with the particle, must be taken into account and can, in fact, significantly increase the effective momentum relaxation time \cite{franosch2011resonances,pesce2014long}.}. Thus it is justified to take the limit $m \to 0$ in SDE~(\ref{eq:mass}). This has to be done carefully and requires a nontrivial calculation. In particular, it is not correct to simply set $m = 0$ and drop the inertial term. 

\begin{widetext}
We will now proceed to outline the derivation of the correct limiting SDE and the corresponding noise-induced drift. We start by rewriting the system (\ref{eq:system}) as
\begin{equation}\label{eq:IIIA1}
dx_t = v_t\,dt = {F(x_t) D(x_t) \over k_{\rm B}T}\,dt + \sqrt{2D(x_t)}\,dW_t - {m \over k_{\rm B}T}D(x_t)\,dv_t \; .
\end{equation}
In integral form, SDE~(\ref{eq:IIIA1}) becomes
\begin{equation}
x_t =  x_0 + \int_0^t{F(x_s) D(x_s) \over k_{\rm B}T}\,ds + \int_0^t\sqrt{2D(x_s)}\,dW_s - \int_0^t{m \over k_{\rm B}T}D(x_s)\,dv_s \; ,
\end{equation}
where the first integral term on the right-hand side is the contribution due to the deterministic (external) forces and the next term is an It\^o integral. In order to derive the noise-induced drift, we will study the limiting behavior  for $m \rightarrow 0$ of the last term. We start by integrating it by parts, obtaining
\begin{equation}\label{eq:ibp}
-\int_0^t{m \over k_{\rm B}T}D(x_s)\,dv_s = {1 \over k_{\rm B}T}[mv_0D(x_0) - mv_tD(x_t)] + \int_0^t{m \over k_{\rm B}T} {dD(x_s)\over dx} v_s^2\,ds \; .
\end{equation}
\end{widetext}
Using Eq.~(\ref{eq:maxwellian}), the boundary terms in Eq.~(\ref{eq:ibp}) go to zero with $m \rightarrow 0$ and, replacing the kinetic energy ($mv_s^2$) by its average ($k_{\rm B}T$) in the last integral, we obtain the effective SDE
\begin{equation}\label{eq:mass:sd}
dx_t = {F(x_t) D(x_t) \over k_{\rm B}T}\,dt \underbrace{+ \frac{dD(x_t)}{dx}\,dt}_{\mbox{spurious drift}} +  \sqrt{2D(x_t)}\,dW_t \; .
\end{equation}
We emphasize that the averaging of the kinetic energy is far from trivial and needs a careful justification; a sketch of the argument (in a more general case) will be given in Section~\ref{sec:case:fdr}. The physical picture is that the velocity $v_t$ is a {\it fast} variable that {\it homogenizes} in the $m \to 0$ limit.  The term {\it adiabatic elimination} is also used in literature to describe this phenomenon.  The above result was proven rigorously in \citet{hottovy2015smoluchowski}.  Related results were obtained earlier in \citet{hanggi1982nonlinear} and in \citet{sancho1982adiabatic}.

\begin{figure}[b]
\includegraphics*[width=3.25in]{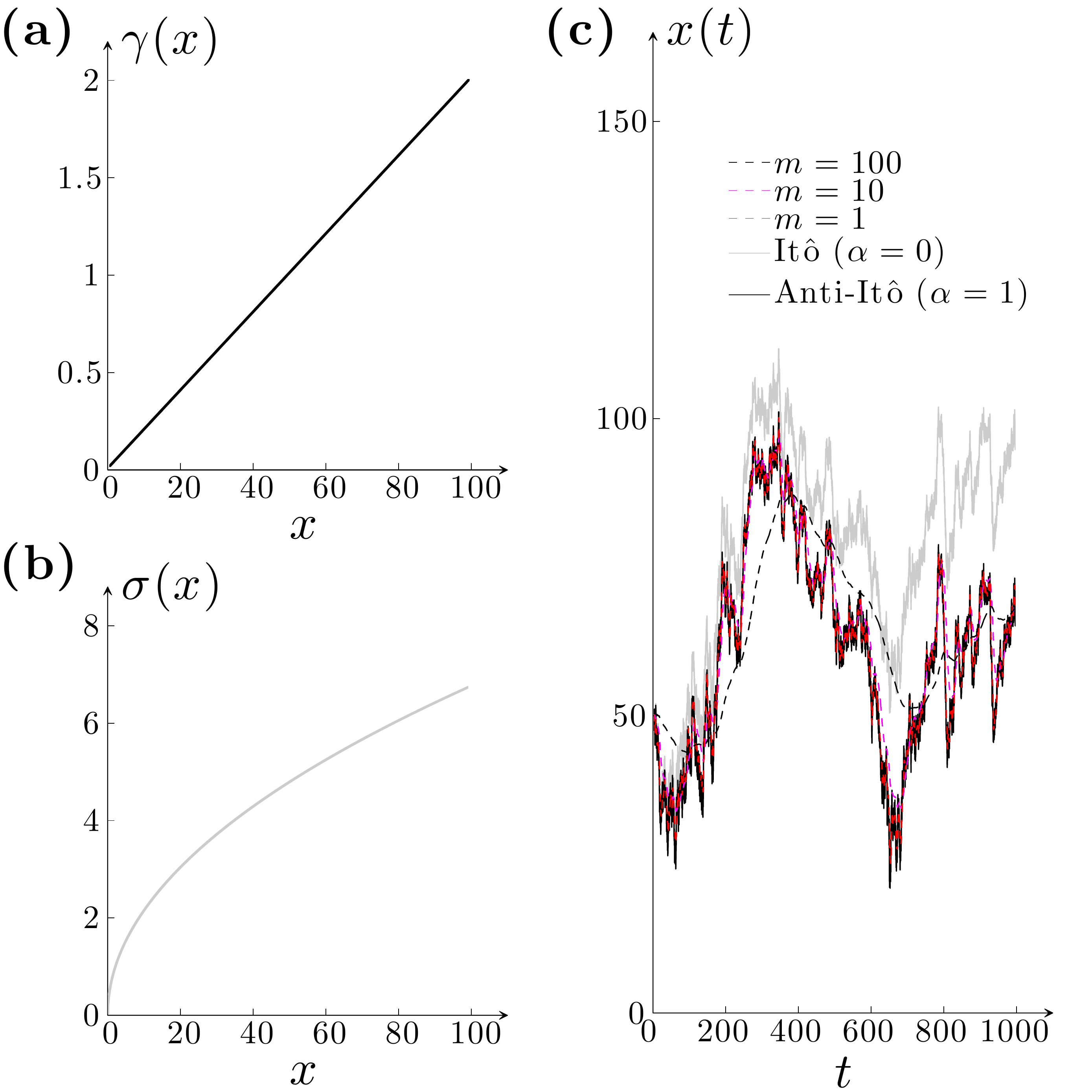}
\caption{(Color online) Limiting SDE for a system satisfying the fluctuation-dissipation relation.
For a Brownian particle in thermal equilibrium (a) $\gamma(x)$ and (b) $D(x)$ are related by the Einstein fluctuation-dissipation relation (Eq.~(\ref{eq:einstein})).
(c) The solutions of the Newton's equations (SDE~(\ref{eq:mass})) for $m \rightarrow 0$ (dashed lines) converge to the solution of the limiting SDE~(\ref{eq:mass:sd}), including the spurious drift (black solid line). The solution without spurious drift (grey solid line) is given for comparison. All solutions are numerically calculated using the same realization of the Wiener process and with $F(x)\equiv 0$.
}
\label{fig5}
\end{figure}

The numerical simulations shown in Fig.~\ref{fig5} give us some insight into the derivation of the limiting SDE and the emergence of the noise-induced drift. We simulate a Brownian particle at equilibrium with a thermal bath, so that its $\gamma(x)$ (Fig.~\ref{fig5}a) and $D(x)$ (Fig.~\ref{fig5}b) are related by the Einstein fluctuation-dissipation relation (Eq.~(\ref{eq:einstein})). The dashed lines in Fig.~\ref{fig5}c represent solutions of SDE~(\ref{eq:mass}) obtained for decreasing values of $m$, but with the same realization of the driving Wiener process. These solutions become  rougher as $m$ decreases and converge towards the solution of the limiting SDE~(\ref{eq:mass:sd}) (black solid line in Fig.~\ref{fig5}c), again calculated using the same realization of the Wiener process. We see that omitting the spurious drift leads to clear deviations, which diverge as a function of time (grey solid line in Fig.~\ref{fig5}c). 

The noise-induced drift in SDE~(\ref{eq:mass:sd}) has been directly observed in at least two sets of experiments. Before proceeding further, we note that, in general, the diffusion $D(x)$ and (total) drift $C(x)$ of an experimental system can be obtained from an experimental discrete time-series $\{x_1,\, ...\, ,\, x_{N}\}$ sampling the system's state at regular intervals $\Delta t$ as the conditional averages
\begin{equation}\label{eq:exp:diffusion}
D(x) = \frac{1}{2 \Delta t} \left< (x_{n+1} - x_n)^2 \mid x_n \cong x \right>
\end{equation}
and
\begin{equation}\label{eq:exp:drift}
C(x) = \frac{1}{\Delta t} \left< x_{n+1} - x_n \mid x_n \cong x \right> \; .
\end{equation}
In experiments $\Delta t$ should be much smaller than the characteristic relaxation time of the system, which is determined by the drift part of the SDE and is typically several orders of magnitude larger than $\tau_{\rm m}$. Furthermore, in the limit $\Delta t \rightarrow 0$ inertial effects come into play \cite{li2013brownian} and, therefore, in practice Eqs.~(\ref{eq:exp:diffusion}) and (\ref{eq:exp:drift}) should only be used in the overdamped limit, i.e., for $\Delta t \gg \tau_{\rm m}$. Similar considerations hold also for other microscopic dynamics determining the evolution of the system, i.e., $\Delta t$ should be much longer than the characteristic times of the dynamics that have been homogenized in the effective SDE.

\begin{figure}[h!]
\includegraphics*[width=3.25in]{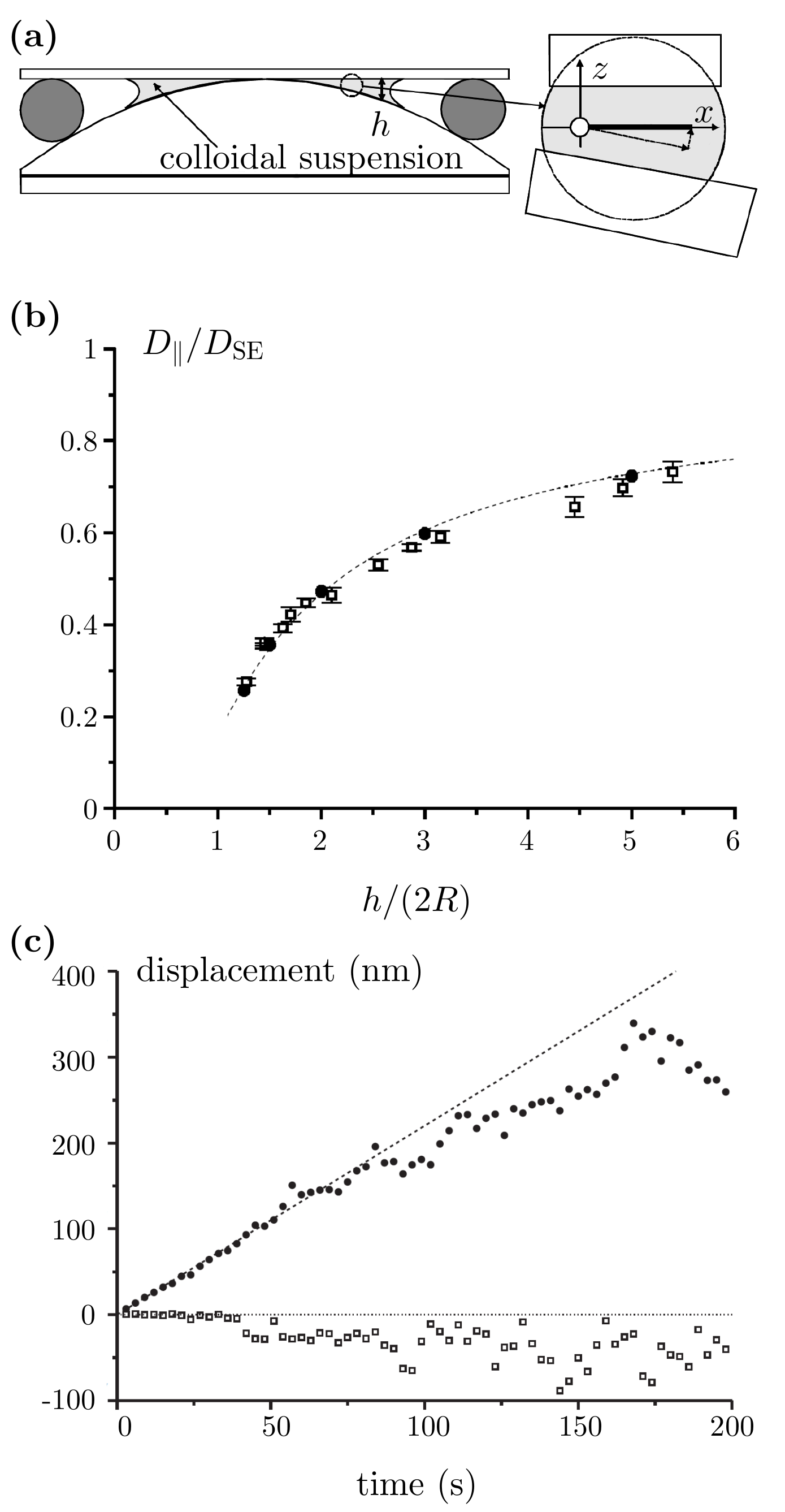}
\caption{Drift without flux.
(a) Cross-section of a sample cell where a colloidal suspension (particle radius $R=1\,{\rm \mu m}$) is confined between a spherical lens and a flat disk, separated by an elastic O-ring. The round inset identifies the observation frame. The height of the cell is denoted by $h$.
(b) Diffusion coefficient $D_{\|}$ normalized to the bilk diffusion coefficient $D_{\|}(\infty) = D_{\rm SE}$ as a function of the relative confinement $h/(2R)$. Open squares are the experimental data; the dotted line is the best fit to the black dots, which are calculated by the collocation method.
(c) Drift of the walkers as a function of time along the diffusion gradient (black dots) and perpendicular to the diffusion gradient (open squares).
Adapted from \citet{lanccon2001drift}.}
\label{fig6}
\end{figure}

The first direct experimental observation of the noise-induced drift was performed by \citet{lanccon2001drift} who studied the Brownian motion of particles trapped between two nearly parallel walls. The experimental sample was realized by placing a droplet of colloidal suspension between a spherical lens (with curvature $L$) and a flat disk, as shown in Fig.~\ref{fig6}a. The spacing $h$ between the flat and curved walls depended on the distance $r$ from the center of the cell as $h = r^2/(2L)$. The colloidal solution consisted of polystyrene spheres (radius $R = 1\,{\rm \mu m}$) suspended in a mixture of ${\rm H_2O}$ and ${\rm D_2O}$ adjusted to cancel any sedimentation effects. The horizontal Brownian motion of the particles was observed using digital video microscopy. The experimental values of the ratio between the measured diffusion coefficient parallel to the walls $D_{\|}(h)$ and the bulk diffusion coefficient $D_{\rm SE}$ were inferred from the measured trajectories using Eq.~(\ref{eq:exp:diffusion}) and are shown in Fig.~\ref{fig6}b (white squares). For the measurement of the noise-induced drift, they fixed the center of the observation frame at a position with $y = 0$ and $x = 300\,{\rm \mu m}$ (inset in Fig.~\ref{fig6}a), corresponding to an average relative confinement $h/(2R) = 1.5$ so that all particles present in the frame were outside of the excluded volume (i.e., $h \le 2R$) and had a diffusion coefficient with the largest $x$-dependence, but no $y$-dependence (to first order). The drift of the Brownian particles over a period of about three minutes is shown in Fig.~\ref{fig6}c. Importantly, they observed no flux and no concentration gradient over a period of a week or more, which is consistent with the (uniform) Boltzmann distribution expected in the absence of external forces and in thermal equilibrium (Eq.~(\ref{eq:BG})).

\begin{figure}[h!]
\includegraphics*[width=3.25in]{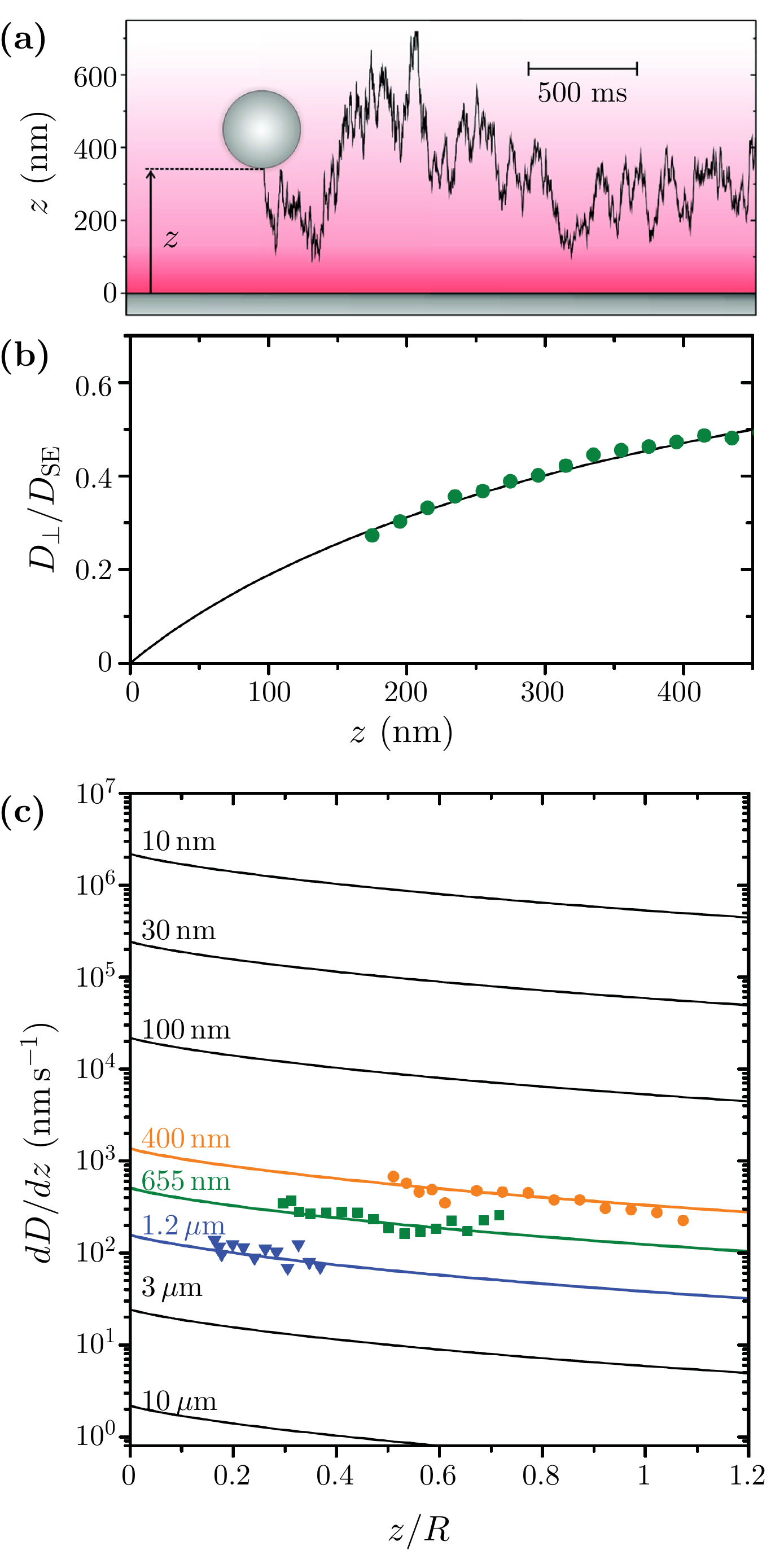}
\caption{(Color online) Experimental measurement of spurious drifts.
(a) A Brownian particle (drawn not to scale) diffuses above a wall in the presence of gravitational and electrostatic forces. Its trajectory's component in the direction perpendicular to the wall is measured with total internal microscopy (TIRM).
Adapted from \citet{volpe2010influence}.
(b) Comparison of measured (symbols) and calculated (line) normalized vertical diffusion coefficient $D_{\perp}(z)/D_{\rm SE}$ for an $R = 400\,{\rm nm}$ particle as a function of the particle-wall separation $z$.
(c) Distance dependence of the theoretically calculated spurious drift ${dD(z) \over dz}$ for various particle radii $R$ (lines). Experimentally measured spurious forces are shown for $R = 400\,{\rm nm}$ (circles), $R = 655\,{\rm nm}$ (squares) and $R = 1180\,{\rm nm}$ (triangles).
Adapted from \citet{brettschneider2011force}.
}
\label{fig7}
\end{figure}

In \citet{volpe2010influence} and \citet{brettschneider2011force}, we studied the Brownian motion of a colloidal particle in water with a diffusion gradient imposed by the presence of the bottom wall of the sample cell, as shown in Fig.~\ref{fig7}a. The external forces acting on the particle were gravity and electrostatic repulsion from the bottom of the sample cell. Since both are vertical, one can separate the horizontal degrees of freedom and write the Newton's equation of motion for the vertical coordinate only, which we will call $z$.  As we have already seen, $D_{\perp}(z)$ decreases near the bottom of the sample cell and its precise form can be found in \citet{brenner1961slow}. The trajectory of a particle close to the wall was measured with total internal reflection microscopy (TIRM), which is a technique that permits one to measure the position of a colloidal particle above a surface with nanometer resolution \cite{volpe2009novel}. From the measured trajectories we obtained $D_{\perp}(z)$ using Eq.~(\ref{eq:exp:diffusion}) (symbols in Fig.~\ref{fig7}b), which is in a very good agreement with the theoretical prediction \cite{brenner1961slow} (line in Fig.~\ref{fig7}b). We were then able to directly measure the spurious drift for particles of various sizes, as shown in Fig.~\ref{fig7}c.

We conclude this section with a brief discussion of how the presence of a noise-induced drift plays a crucial role within the context of the measurement of forces acting on Brownian particles in a liquid medium. The forces acting on a microscopic object immersed in a liquid medium can be assessed either by studying the underlying potential or by studying their effect on the object's trajectory \cite{volpe2010influence,brettschneider2011force}. The first approach --- to which we shall refer as {\it equilibrium distribution method} --- requires sampling of the equilibrium distribution $\rho(x)$ of the particle position (see also Eq.~(\ref{eq:BG})). The force can then be derived as
\begin{equation}
F(x) = -\frac{dU(x)}{dx} = \frac{k_{\rm B}T}{\rho(x)} \frac{d\rho(x)}{dx} \; .
\end{equation}
This method can only be applied under conditions where the investigated system is in thermodynamic equilibrium with a heat bath. The second method --- to which we shall refer as {\rm drift method} --- does not require the object to be in (or even close to) thermal equilibrium. This method requires the measurement of $D(x)$ and $C(x)$ from experimentally obtained trajectories, including a correction for the presence of a spurious (noise-induced) force. The force can then be calculated as
\begin{equation}
F(x) = \gamma(x) C(x) \underbrace{- \gamma(x) \frac{dD(x)}{dx}}_{\mbox{spurious force}} \; .
\end{equation}
This method has the advantage that it can be applied also to systems that are intrinsically out-of-equilibrium, e.g., molecular machines, transport through pores, DNA stretching; however, it requires recording of the object's trajectory with high sampling rates, which can be technologically challenging, in particular when combined with a high spatial resolution.

\subsection{Diffusive systems not satisfying the fluctuation-dissipation relation}\label{sec:case:fdr}

\begin{figure}
\includegraphics*[width=3.25in]{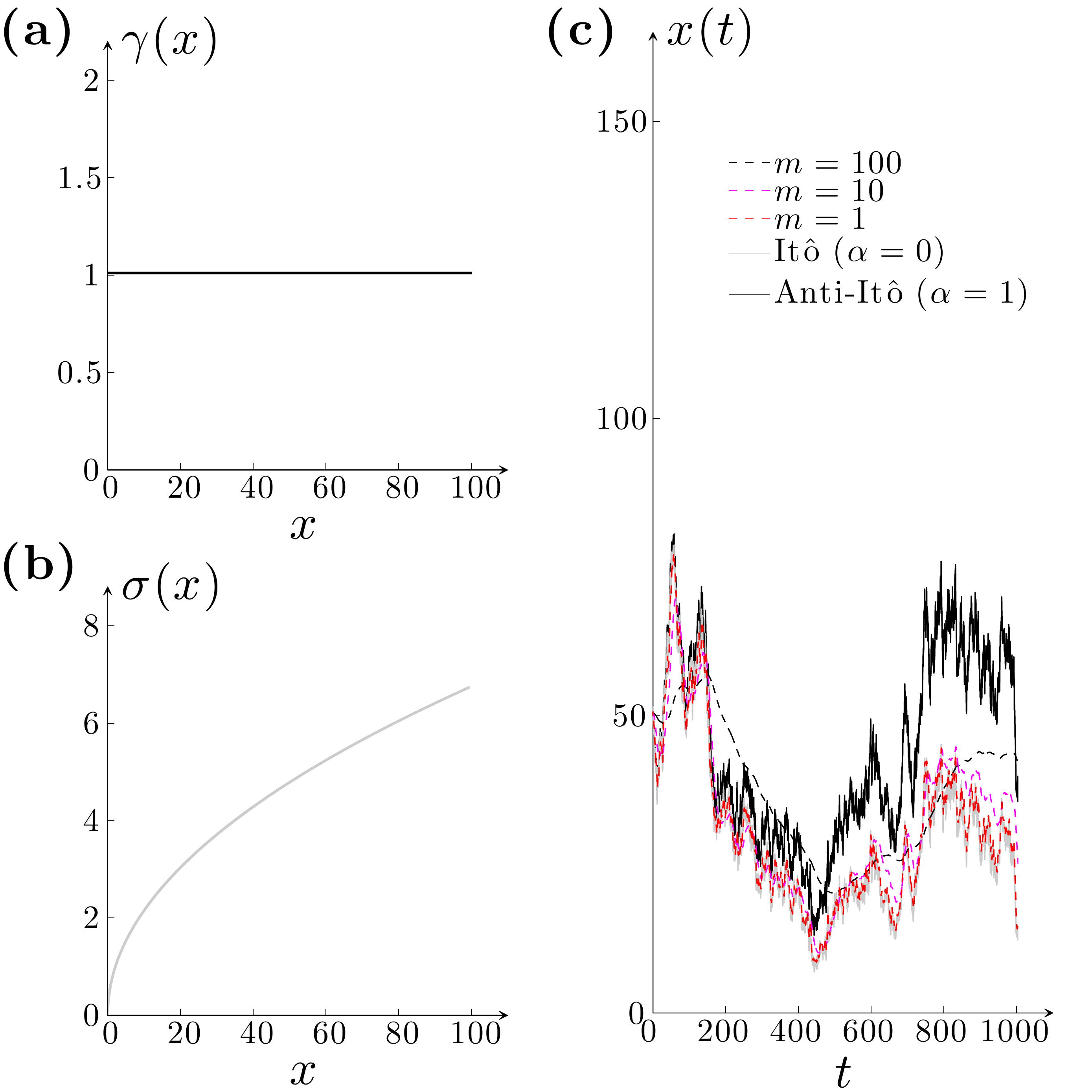}
\caption{(Color online) Limiting SDE for a system not satisfying the fluctuation-dissipation relation.
Consider a system for which (a) $\gamma(x) \equiv \mbox{constant}$ and (b) $\sigma(x)$ is state-dependent. (c) The solutions of SDE~(\ref{eq:nofdr}) for $m\to0$ (dashed lines) converge to the solution of the approximated SDE~(\ref{eq:B}), which in this case corresponds to the It\^o interpretation (grey solid line) of the equation $dx_t = \frac{F(x_t)}{\gamma}\,dt + \frac{\sigma(x_t)}{\gamma}\,dW_t$, as the noise-induced drift equals zero ($\alpha = 0$) when $\gamma$ is constant. The solution for the anti-It\^o integral ($\alpha = 1$, black solid line) is given for comparison. All solutions are obtained for the same realization of the Wiener process.
}
\label{fig8}
\end{figure}

While in Section~\ref{sec:case:bm} we considered systems in thermal equilibrium with a heat bath that satisfy the fluctuation-dissipation relation (Eq.~(\ref{eq:einstein})), in this section we consider the zero-mass limiting behavior of a larger class of models for which $\gamma(x)$ and $\sigma(x)$ are allowed to vary independently from each other. This is a very general class of noisy dynamical systems, with many interesting examples
and applications, see, e.g., \citet{ao2007existence} and \citet{hottovy2012noise}. Using the methods of \citet{hottovy2015smoluchowski}, we will thus study the general SDE
\begin{equation}\label{eq:nofdr}
m\ddot{x}_t = F(x_t) - \gamma(x_t)\dot{x}_t + \sigma(x_t)\eta_t \; ,
\end{equation}
where the damping and diffusion terms are not necessarily related by the  fluctuation-dissipation relation (Eq.~(\ref{eq:einstein})). As we will see, for a wide class of such systems the effective equation in the $m \to 0$ limit is
\begin{equation}\label{eq:B}
dx_t = \Bigg[
{F(x_t) \over \gamma(x_t)} 
\underbrace{ - {\sigma(x_t)^2 \over 2\gamma(x_t)^3}{d\gamma(x_t) \over dx} }_{\mbox{noise-induced drift}}
\Bigg]
\,dt + {\sigma(x_t) \over \gamma(x_t)}\,dW_t \; .
\end{equation}
An example of such a system is illustrated in Fig.~\ref{fig8}: for the case with $\gamma(x) \equiv \mbox{constant}$ (Fig.~\ref{fig8}a) and $\sigma(x)$ state-dependent (Fig.~\ref{fig8}b), the solutions of SDE~(\ref{eq:nofdr}) for $m\to0$ (dashed lines in Fig.~\ref{fig8}c) converge to the solution of the approximate SDE~(\ref{eq:B}) (grey solid line in Fig.~\ref{fig8}c); note that in this case the noise-induced drift is zero.

In general, there is no relation between noise and damping coefficients if the noise is {\it external}, as in an electrical circuit driven by a noise source.  Such a circuit with a colored noise and involving a delayed response is studied in Section~\ref{sec:case:color}. Another physical example
described by an equation of the form of SDE~(\ref{eq:nofdr}) is diffusion of a Brownian particle in a temperature gradient. This system shows more interesting phenomena when it is driven by a colored noise; a simple model of this type is studied in \citet{hottovy2012thermophoresis}. Brownian motion in a diffusion gradient, discussed in Section~\ref{sec:case:bm} is yet another special case of a system described by the SDE~(\ref{eq:nofdr}) and the result outlined there is a special case of SDE~(\ref{eq:B}). 

To analyze the general SDE~(\ref{eq:nofdr}), we follow the argument in \citet{hottovy2015smoluchowski}, concentrating on the main steps and leaving out technical details and estimates. The first step in the derivation of the limiting equation is the same as in the special case studied in Section~\ref{sec:case:bm}:  introducing the velocity $v_t$, we rewrite SDE~(\ref{eq:B}) as
\begin{equation}\label{eq: rewrite}
dx_t = v_t\,dt = {F(x_t) \over \gamma(x_t)}\,dt + {\sigma(x_t) \over \gamma(x_t)}\,dW_t - {m \over \gamma(x_t)}\,dv_t \; .
\end{equation}
The first two terms on the right-hand side do not depend explicitly on $m$ and thus remain unchanged in the limit $m \to 0$.  To derive the limiting contribution of the third term, we use the product rule:
\begin{equation}
{m\over\gamma(x_t)}\,dv_t = m\,d({v_t\gamma(x_t)}) - d\left({m\over\gamma(x_t)}\right)v_t \; .
\end{equation}
While the equipartition theorem no longer holds in this generality, we will show that the fast oscillations of the velocity allow to replace in the integrals the expression $mv_t^2\,dt$ by a function of $x_t$ (homogenization or adiabatic elimination of the fast variable $v_t$ \cite{pavliotis2008multiscale}).  This is done by first showing that $mv_t$ converges to zero as $m \to 0$ (a technical step, involving careful estimates \cite{hottovy2015smoluchowski}).  It follows that in the integral form of the last equation, i.e.,
\begin{equation}\label{eq: integral}
\int_0^t {m\over\gamma(x_s)}\,dv_s = m\left[{v_t\over\gamma(x_t)} - {v_0\over\gamma(x_0)}\right] - \int_0^t d\left({m\over\gamma(x_s)}\right)v_s \; ,
\end{equation}
the first term on the right-hand side vanishes in the limit.  The integrand in the second term equals $\gamma(x_t)^{-2} \, {d \over dx}\gamma(x_t) \, mv_t^2$.  To find its homogenization limit, we study the expression $d[(mv_t)^2]$.  On the one hand, this quantity becomes zero when $m \to 0$.  On the other hand, using the It\^o product formula \cite{oksendal2013stochastic}, we have
\begin{equation}
d(mv_t \cdot mv_t) = 2mv_t\,d(mv_t) + [\,d(mv_t)]^2 \; .
\end{equation}
Substituting for $m\,dv_t$ the expression on the right-hand side of SDE~(\ref{eq:nofdr}), we obtain
\begin{equation}
d[(mv_t)^2] = 2mv_t[F(x_t)\,dt - \gamma(x_t)v_t\,dt + \sigma(x_t)\,dW_t] + \sigma(x_t)^2\,dt.
\end{equation}
The first and third terms on the right-hand side converge to zero, since $mv_t$ does (the rigorous argument again requires some care \cite{hottovy2015smoluchowski}).  
Since in the limit $m \to 0$, the whole expression converges to zero, it follows that  $mv_t^2\,dt$ is asymptotically equivalent to ${\sigma(x_t)^2 \over 2\gamma(x_t)}\,dt$.  We thus obtain from Eq.~(\ref{eq: integral})
\begin{equation}
\int_0^t {m\over\gamma(x_s)}\,dv_s \to \int_0^t{\sigma(x_t)^2 \over 2\gamma(x_t)^3} {d\gamma(x_t) \over dx} \,dt
\end{equation}
in the zero mass limit.  Substituted into SDE~(\ref{eq: rewrite}), this gives the limiting SDE~(\ref{eq:B}).  

SDE~(\ref{eq:nofdr}) can be generalized to multidimensional (i.e., vector) systems as
\begin{equation}\label{eq:SDEgeneral}
\left\{\begin{array}{ccl}
\displaystyle d\bm{x}_t^m 
& = & 
\displaystyle \bm{v}_t^m\,dt \; , \\[6pt]
\displaystyle d\bm{v}_t^m 
& = & 
\displaystyle \left[ \frac{\bm{F}(\bm{x}_t^m)}{m} - \frac{\bm{\bm{\gamma}}(\bm{x}_t^m)}{m}\bm{v}^m_t \right] \,dt + \frac{\bm{\sigma}(\bm{x}_t^m)}{m}\,d\bm{W}_t \; ,
\end{array}\right.
\end{equation} 
where ${\bf W}$ is a vector Wiener process (i.e., the components of ${\bf W}$ are independent Wiener processes), and $\boldsymbol{\gamma}$ and $\boldsymbol{\sigma}$ are matrices. The method described above can be adapted to derive the limit of SDE~(\ref{eq:SDEgeneral}) as the mass goes to zero.  The main idea is the same as in the one-dimensional case, but calculations become more complicated and the description of the noise-induced drift is more involved:  it is expressed using a unique solution of a matrix equation (the Lyapunov equation).  The final result becomes more implicit, since the solution of the Lyapunov equation is, in general, expressed as an integral over an auxiliary parameter.  In an important class of cases, the Lyapunov equation has an explicit solution and the limiting equation becomes explicit as well.  The precise form of the limiting equation is:
\begin{equation}\label{eq:SKlimit}
d\bm{x}_t=\left[ \bm{\gamma}^{-1}(\bm{x}_t)\bm{F}(\bm{x}_t)+\bm{S}(\bm{x}_t) \right] dt + \bm{\gamma}^{-1}(\bm{x}_t)\bm{\sigma}(\bm{x}_t)d\bm{W}_t,
\end{equation}
Here $\bm{S}(\bm{x}_t)$ is the \emph{noise-induced drift} whose $i^{\rm th}$ component equals
\begin{equation}
\label{eq:spurious}
S_i(\bm{x}) = \sum_{j,l}\frac{\partial}{\partial x_{l}}[(\gamma^{-1})_{ij}(\bm{x})]J_{j{l}}(\bm{x}),
\end{equation}
and $\bm{J}$ is the matrix solving the Lyapunov equation
\begin{equation}\label{eq:Lyapunov}
\bm{J}\bm{\gamma}^* + \bm{\gamma}\bm{J} = \bm{\sigma}\bm{\sigma}^*.
\end{equation}
When all eigenvalues of $\bm{\gamma}$ have negative real parts, the solution is given by the formula \cite[Chapter 11]{bellman1970introduction}
\begin{equation}
{\bf J} = -\int_0^{\infty}e^{y\bm{\gamma}} \bm{\sigma}\bm{\sigma}^* e^{y\bm{\gamma}^*} \, dy \; .
\end{equation}
Note that when $\boldsymbol{\gamma} = \boldsymbol{\gamma}^*$ commutes with $\boldsymbol{\sigma}\boldsymbol{\sigma}^*$, the solution of the Lyapunov equation is explicitly given by $\boldsymbol{J} = {1\over2} \boldsymbol{\gamma}^{-1} \boldsymbol{\sigma}\boldsymbol{\sigma}^*$.

The zero-mass limits of equations similar to SDE~(\ref{eq:SDEgeneral}) have been studied by many authors beginning with Smoluchowski \cite{smoluchowski1916drei} and Kramers \cite{kramers1940brownian}. In the case where $F=0$ and $\gamma$ and $\sigma$ are constant, the solution to SDE~(\ref{eq:nofdr}) converges to the solution of SDE~(\ref{eq:B}) almost surely \cite{nelson1967dynamical}.  \citet{schuss1980theory} treated the case including an external force by entirely different methods. \citet{hanggi1982nonlinear} identified the limit with position-dependent noise and friction for the case when the fluctuation-dissipation relation is satisfied and \citet{sancho1982adiabatic} for the general one-dimensional case (the multidimensional case is also discussed there but without complete proof). \citet{hottovy2012noise} used the homogenization techniques described in \citet{papanicolaou1975introduction}, \citet{schuss1980theory} and \citet{pavliotis2008multiscale} to compute the limiting backward Kolmogorov equation corresponding to Eq.~(\ref{eq:nofdr}) as mass is taken to zero. \citet{pardoux2003poisson} proved rigorously convergence in distribution for equations of the same type as SDE~(\ref{eq:SDEgeneral}), under somewhat stronger assumptions than those made in \citet{hottovy2015smoluchowski}. \citet{freidlin2004some} gave the first rigorous proof of strong convergence in the zero-mass limit for $\bm{\gamma}$ constant and $\bm{\sigma}$ position-dependent. \citet{hottovy2015smoluchowski} provided the first rigorous derivation of the zero-mass limit of SDE~(\ref{eq:SDEgeneral}) for a multidimensional system with general friction and noise coefficients.

The general form of SDE~(\ref{eq:SDEgeneral}) allows to treat many interesting physical situations, including the case when the force $\bf{F}$ is not conservative.  In this case, there is no known explicit formula for the stationary measure of the dynamics defined by SDE~(\ref{eq:SDEgeneral}), even when the system satisfies the fluctuation-dissipation relation.  Nevertheless, the general theorem applies, giving the limiting equation for $x_t$.

As another application of the general scheme given by SDE~(\ref{eq:SDEgeneral}), suppose the white noise in SDE~(\ref{eq:nofdr}) is replaced by a colored (i.e., time-correlated) stationary noise process, which is itself a solution of a stochastic differential system.  For example, $\eta^{\tau}$ may be an Ornstein-Uhlenbeck process:
\begin{equation}\label{eq: OU}
d\eta^{\tau}_t = -{a \over \tau}\eta^{\tau}_t\,dt + {1 \over \tau}\,dW_t \; .
\end{equation}
Defining $\chi_t = \int_0^t\eta^{\tau}_s\,ds$ we introduce a new, compound space variable $(x, \chi)$ and the corresponding velocity $(v, \eta^{\tau})$.  
If the parameter $\tau$ scales linearly with $m$, the variables $(x, \chi)$ and $(v, \eta^{\tau})$ satisfy a system of equations of the same form as SDE~(\ref{eq:SDEgeneral}).  The above general result applies, yielding an effective equation for a system in which the momentum relaxation time $\tau_{\rm m}$ and the characteristic noise correlation time $\tau$ go to zero at the same rate.  The details are given in \citet{hottovy2015smoluchowski}. In Section~\ref{sec:case:color} we will see that $\tau$ can also interact with the feedback delay time of the system.

We emphasize that the induced drift $\bf{S}$ may be nonzero even if the noise coefficient is constant, as opposed to the well-known {\it It\^o-to-Stratonovich correction} \cite{oksendal2013stochastic}.  In fact, while the latter can be presented (using Wong-Zakai theorem \cite{wong1965convergence}) as a special case of noise-induced drift, the noise-induced drift phenomenon is much more general.

\subsection{Delayed multiplicative feedback and colored noise}\label{sec:case:color}

\begin{figure}[b]
\includegraphics*[width=3.25in]{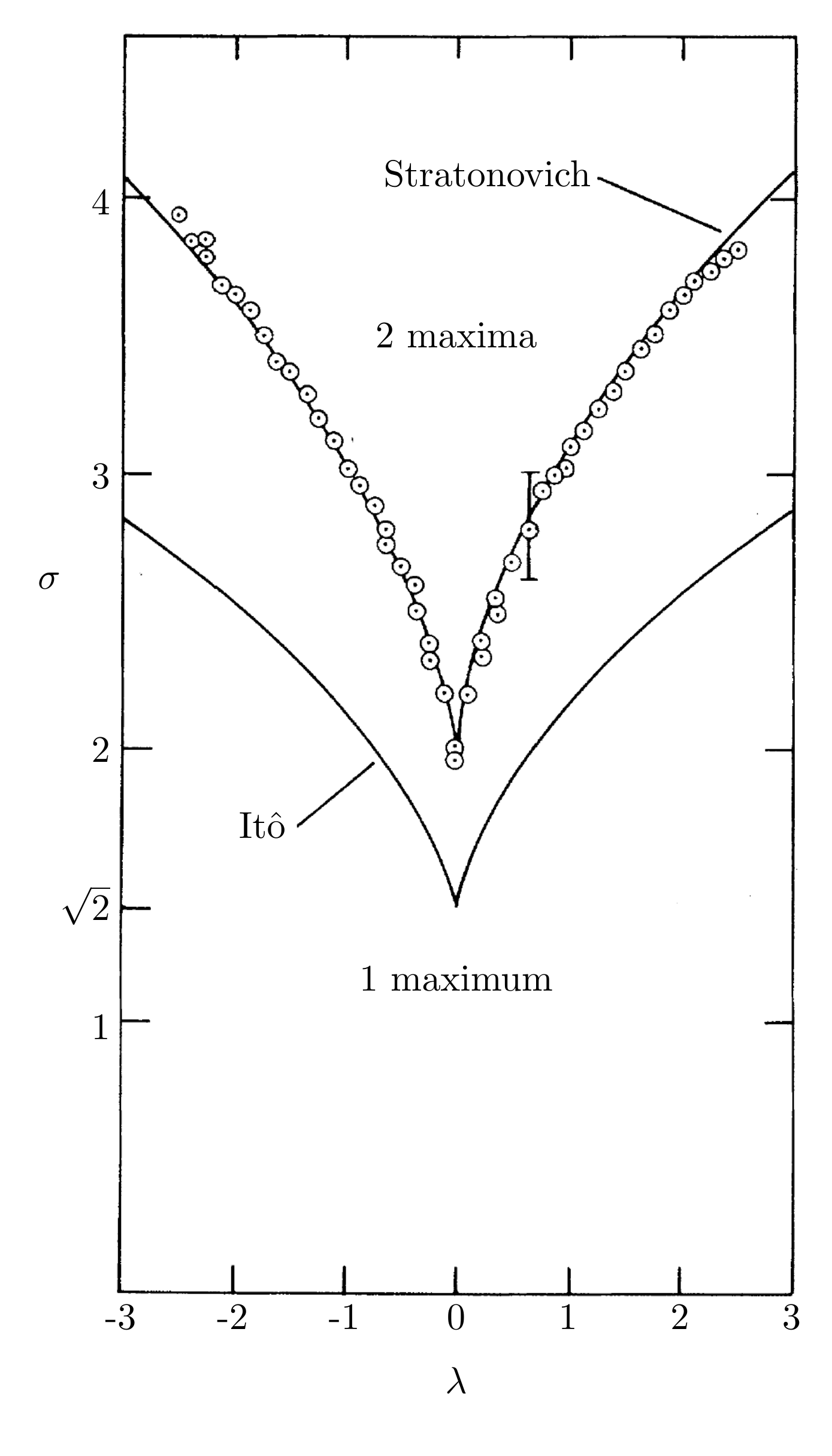}
\caption{Experimental observation of a noise-induced phase transition. Phase diagram of the electric circuit with multiplicative noise used by \citet{smythe1983observation}. Depending on the input noise parameters, namely its mean $\lambda$ and its variance $\sigma$, the circuit could be either monostable (1 maximum) or bistable (2 maxima). The experimentally measured transition between the two regimes (open circles) agrees with the predictions obtained by integrating the SDE describing the circuit according to the Stratonovich convention. The reason for this is that the driving noise is colored. The predictions according to the It\^o convention are shown for comparison.
Adapted from \citet{smythe1983observation}.
}
\label{fig9}
\end{figure}

White noise does not exist in real systems, since its correlation time is strictly equal to zero (and even as a mathematical object it does not have well defined realizations which would be functions of time) \cite{oksendal2013stochastic}. Colored noises, instead, are more regular mathematical objects than the white noise and more similar to signals that can be actually realistically generated. Thus, it is natural to consider SDEs driven by colored noise. An SDE with colored noise can be interpreted as a usual ordinary differential equation for each noise realization. However, the very correlation effects we want to model make such equations harder to study. The characteristic time of the noise correlations, $\tau$, becomes an important time scale of the model, whose properties often simplify in the limit when $\tau \to 0$. Such limit is studied in the classical work of \citet{wong1965convergence}, who consider a sequence of SDEs driven by colored noises with symmetric covariance functions and with correlation times $\tau_n \to 0$ and show that their solutions converge to the solution of the corresponding Stratonovich equation driven by the white noise. A more general result can be found in \citet{kurtz1991weak}; see also \citet{kupferman2004ito}, where such limits are studied using homogenization methods, and \citet{freidlin2011smoluchowski}. We remark that all these results can be recovered by the methods discussed in Section~\ref{sec:case:fdr}.

A system obeying an SDE with a colored noise was experimentally realized by \citet{smythe1983observation} as an eletrical circuit driven by a multiplicative noisy voltage input. Depending on the mean and variance of the noise, the output voltage of the circuit could have a probability density with either one or two maxima, and the precise form of the phase diagram depended on whether the equation describing the circuit was interpreted using the It\^o or Stratonovich integral. As shown in Fig.~\ref{fig9}, the results of \citet{smythe1983observation} were in quite good agreement with the theoretical predictions based on the Stratonovich interpretation, illustrating the role of the colored noise, as mathematically described by the Wong-Zakai theorem.

\begin{figure*}
\includegraphics*[width=6.5in]{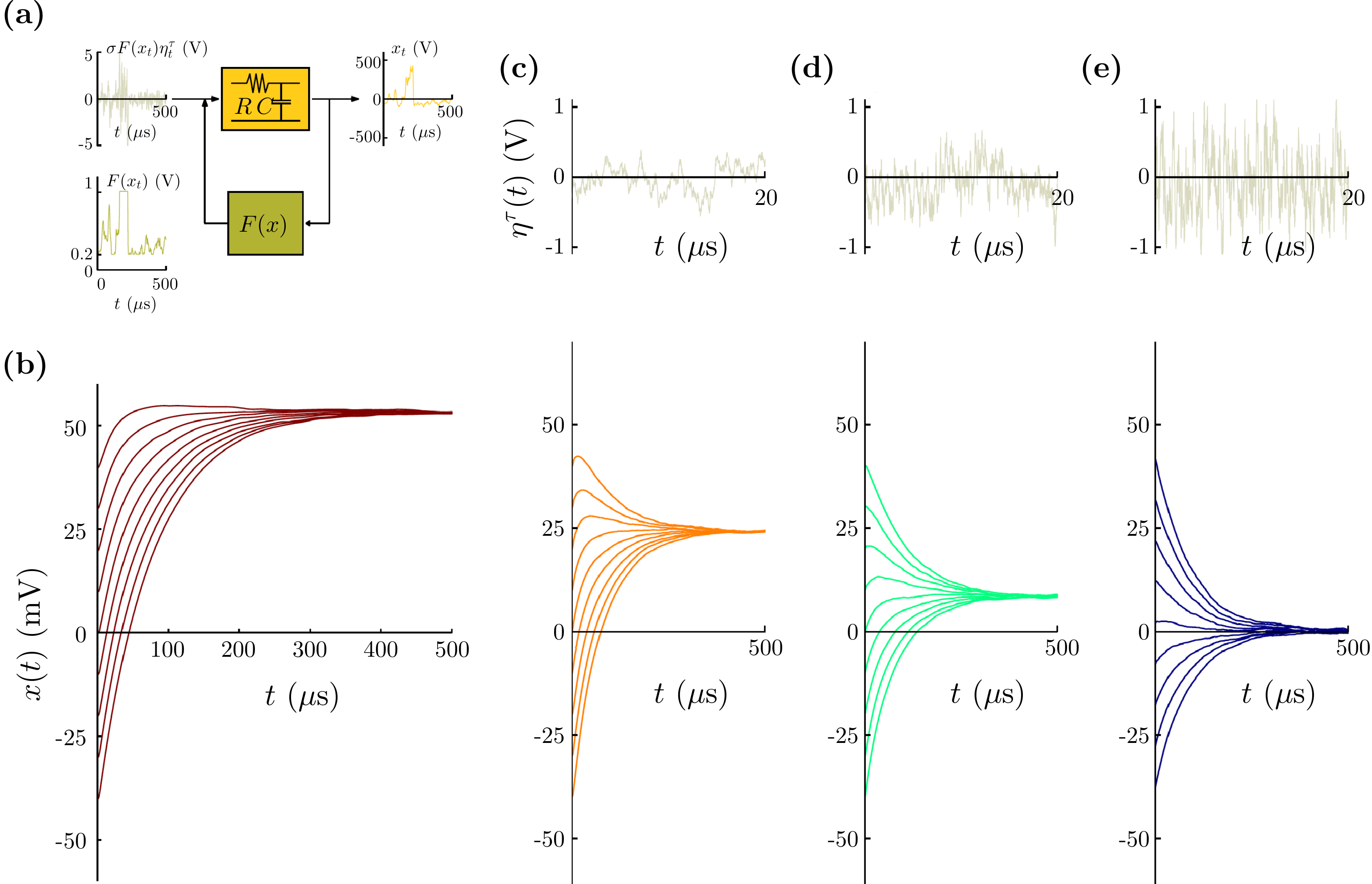}
\caption{Stochastic dynamical system driven by multiplicative noise with delayed feedback.
(a) Schematic representation of a stochastic dynamical system (an electric circuit) with multiplicative feedback $F(x)$: the driving colored noise $\eta^{\tau}_t$ ($\tau = 1.1\,{\rm ms}$) is multiplied by a function of the system's state $x_t$.
(b) Average of $1000$ trajectories for various initial conditions. These results are in agreement with the Stratonovich treatment of the circuit SDE.
(c-e) Samples of input noises $\eta^{\tau}_t$ (top) and average of $1000$ trajectories (bottom) for various initial conditions with $\tau = 0.6$, $0.2$ and $0.1\,{\rm ms}$ respectively for (c), (d) and (e). From (b) to (e), there is a shift of the equilibrium towards $x=0$, corresponding to a crossover from the Stratonovich solution to the It\^o solution of the circuit SDE.
Reproduced from \citet{pesce2013stratonovich}.
}
\label{fig10}
\end{figure*}

\begin{figure*}[t!]
\includegraphics*[width=4in]{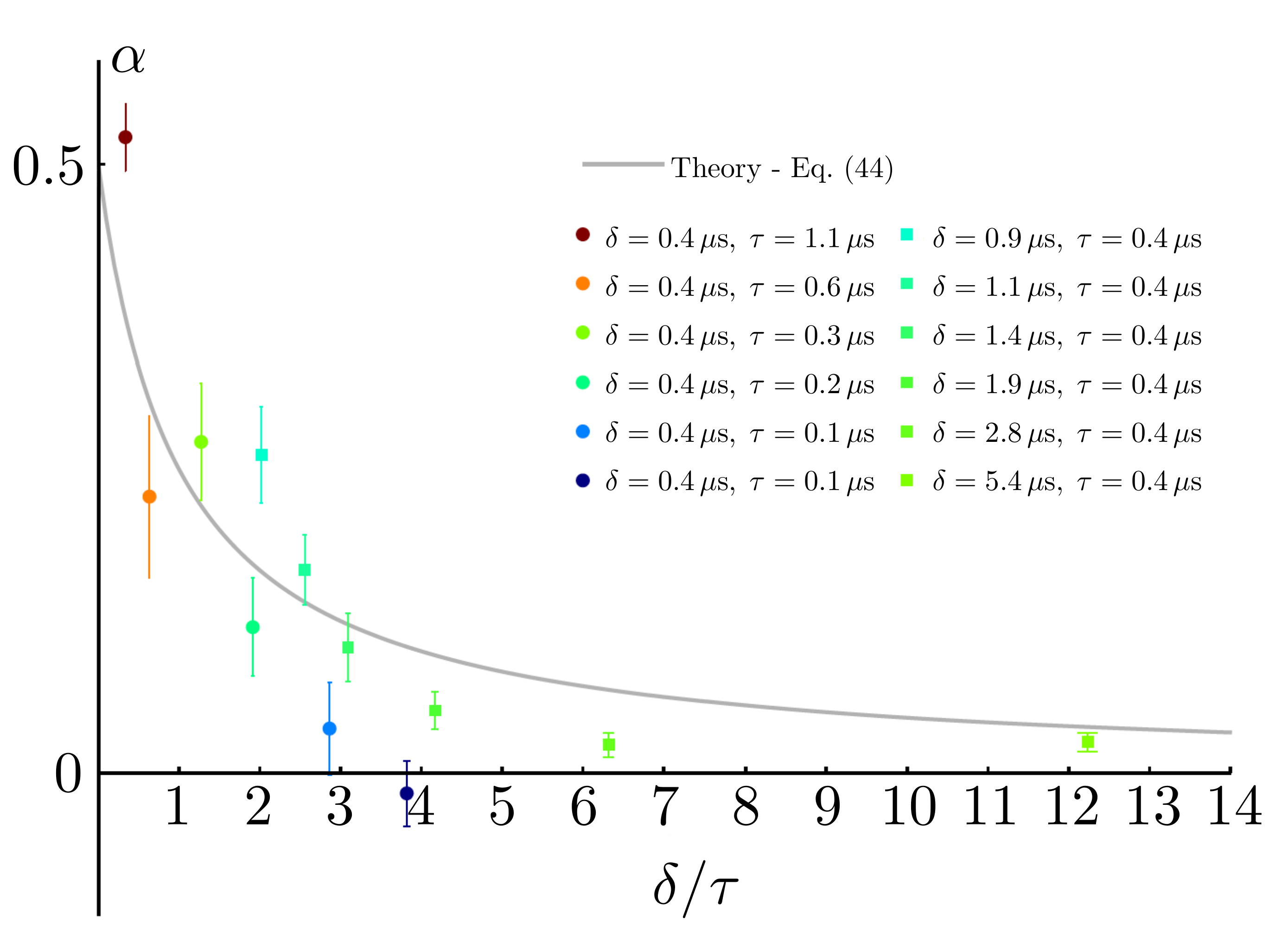}
\caption{(Color online) Dependence of $\alpha$ on $\delta/\tau$. 
$\alpha$ varies from 0.5 (Stratonovich integral) to 0 (It\^o integral) as $\delta/\tau$ increases. The solid line represents theoretical results (Eq.~(\ref{eq:mathe_alpha})); the dots represent the experimental values of $\alpha$ for fixed $\delta = 0.4\,\mathrm{\mu s}$ and varying $\tau$; and the squares the experimental values for fixed $\tau = 0.4\,\mathrm{\mu s}$ and varying $\delta$. The error bars represent one standard deviation obtained by repeating the experimental determination of the ratio $\delta/\tau$ ten times.
Adapted from \citet{pesce2013stratonovich}.
}
\label{fig11}
\end{figure*}

We will now consider in detail the experiment performed by \citet{pesce2013stratonovich} using an RC electric circuit driven by a multiplicative colored noise (Fig.~\ref{fig10}a), in which the output voltage was fed back into the system and multiplicatively coupled to the noise source, after going through a nonlinear filter. Unlike the circuit studied by \citet{smythe1983observation}, the circuit studied by \citet{pesce2013stratonovich} involved a delay in the feedback cycle. The SDE describing the evolution of the voltage in the circuit presented in Fig.~\ref{fig10}a is
\begin{equation}\label{eq:pesce}
dx_t = -kx_t\,dt + \sigma F(x_{t-\delta})\eta^{\tau}_t\,dt \; ,
\end{equation}
where $k = {1 \over RC}$, $R$ is the resistance of the circuit, $C$ is its conductance, and $F$ represents the modulation by the filter. The colored noise $\eta^{\tau}$ is an Ornstein-Uhlenbeck process with mean zero and with the characteristic time of correlation decay equal $\tau$ (i.e., the stationary solution of SDE~(\ref{eq: OU}) with $a=1$).  $\delta$ is the time delay resulting from the application of the filter and $\sigma$ denotes the (constant) noise intensity.

We studied SDE~(\ref{eq:pesce}) in the limit of small $\tau$ and $\delta$.  Mathematically, this meant making $\tau$ and $\delta$ proportional to a small parameter $\epsilon$ and taking the limit $\epsilon \to 0$, keeping the ratio ${\delta \over \tau}$ constant.  The limiting SDE turned out to be
\begin{equation}\label{eq:circuitlim}
dx_t  = \Bigg[-kx_t \underbrace{+ \alpha \sigma^2F(x_t){dF(x_t) \over dx}}_{\mbox{noise-induced drift}}\Bigg]\,dt + \sigma F(x_t)\,dW_t \; .
\end{equation}
The second term has the same structure as the noise-induced drift in the Brownian motion case:  it is proportional to the product of the original noise coefficient $\sigma F(x)$ and its spatial derivative.  The proportionality constant depends on the time scales of the problem as 
\begin{equation}\label{eq:mathe_alpha}
\alpha = {0.5 \over 1 + {\delta \over \tau}} \; ,
\end{equation}
which agrees well with the experimental results, as shown in Fig.~\ref{fig11}. Varying $\delta$ and $\tau$ one can interpolate continuously between $\alpha = 0.5$ and $\alpha = 0$ which can be viewed as a crossover of the system from Stratonovich to It\^o behavior of the system. SDE~(\ref{eq:circuitlim}) is written here in the It\^o form, but it can be interpreted according to another convention, corresponding to another choice of the parameter $\alpha$, as described in Section~\ref{sec:simple}. In this language, the presence of such delay made the SDE describing the behavior of the electric circuit with multiplicative noise cross over from obeying the Stratonovich convention ($\alpha=0.5$) to obeying the It\^o convention ($\alpha = 0$), as the ratio between the colored noise correlation time $\tau$ and the feedback delay $\delta$ varied (Eq.~(\ref{eq:mathe_alpha})), as shown in Figs.~\ref{fig10}b-e. The fact that this transition occurs as $\tau$ becomes close to $\delta$, i.e., $\delta/\tau \approx 1$ (Fig.~\ref{fig11}), can be qualitatively explained as follow: if $\delta = 0$, there is a correlation between the sign of the input noise and the time-derivative of the feedback signal, which incidentally is the underlying reason why the process converges to the Stratonovich solution \cite{wong1965convergence}; however, if $\delta \gg \tau$, this correlation disappears, effectively randomizing the time-derivative of the feedback signal with respect to the sign of the input noise and leading to a situation where the system loses its memory. While this crossover between two stochastic integration conventions was emphasized in \citet{pesce2013stratonovich}, we remark here that this is just a possible way of interpreting the noise-induced drift.

We now discuss the mathematical derivation in more detail, including the approximation which makes the mathematical analysis possible. To derive the limiting SDE~(\ref{eq:circuitlim}), we approximate the delay equation by an SDE without delay and apply the method based on integration by parts, outlined in Section~\ref{sec:case:fdr}.  First, we define a time-shifted process
\begin{equation}
z_t = x_{t-\delta}
\end{equation}
and rewrite the equation as
\begin{equation}
dz_{t+\delta} = -kz_{t+\delta} + \sigma F(z_t)\eta^{\tau}_t\,dt \, .
\end{equation}
Introducing the process 
\begin{equation}
v_t = {dz_t \over dt},
\end{equation}
we use approximations
\begin{equation}
z_{t+\delta} \approx z_t + \delta v_t
\end{equation}
and, accordingly,
\begin{equation}
dz_{t+\delta} \approx dz_t + \delta\,dv_t \, .
\end{equation}
Substituting these expressions into SDE~(\ref{eq:pesce}) and solving for $dv_t$, we obtain the system
\begin{equation}
\left\{\begin{array}{ccl}
\displaystyle dz_t 
& = & 
\displaystyle v_t\,dt \; ,
 \\[6pt]
\displaystyle  dv_t 
& = & 
\displaystyle -{1 \over \delta}k z_t\,dt - \left({1 \over \delta} + k\right)v_t\,dt + {1 \over \delta}\eta^{\tau}_t\,dt \; .
\end{array}\right.
\end{equation}
This system can be studied using the method outlined in Section~\ref{sec:case:fdr}:  we add the equations describing the process $\eta^{\tau}$ to the system and apply the general method of \citet{hottovy2015smoluchowski} to identify the limiting system which matches the experimental results.

In \citet{mcdaniel2014sde}, the same method is applied to a much more general system of delayed SDE driven by several colored noises with couplings that are functions of the delayed dynamical variables:
\begin{equation} \label{eq: SDDE}
d \bm{x} _t = \bm{f}(\bm{x} _t) dt + \bm{g}(\bm{x}_{t - \delta}) \bm{\eta}^{\tau}_t dt \, ,
\end{equation}
where $\bm{x} _t = (x^1 _t, ..., x^i _t, ..., x^m _t)^{\rm T}$ is the state vector (the superscript ``${\rm T}$" denotes transposition), $\bm{f}(\bm{x} _t) = (f^1 (\bm{x}_t), ..., f^i (\bm{x}_t), ..., f^m (\bm{x}_t))^{\rm T}$ is a vector-valued function describing the deterministic part of the dynamical system, 
\begin{equation}
\bm{g}(\bm{x}_{t - \delta}) =
\left[\begin{array}{ccccc}
g^{11}(\bm{x}_{t - \delta}) & \dots & g^{1j}(\bm{x}_{t - \delta}) & \dots & g^{1n}(\bm{x}_{t - \delta})  \\
\vdots & \ddots & \vdots & \ddots & \vdots \\
g^{i1}(\bm{x}_{t - \delta}) & \dots & g^{ij}(\bm{x}_{t - \delta}) & \dots & g^{in}(\bm{x}_{t - \delta}) \\
\vdots & \ddots & \vdots & \ddots & \vdots \\
g^{m1}(\bm{x}_{t - \delta}) & \dots &  g^{mj}(\bm{x}_{t - \delta}) & \dots & g^{mn}(\bm{x}_{t - \delta})
\end{array}\right]
\end{equation} is a matrix-valued function, $\bm{x} _{t - \delta} = (x^1 _{t - \delta _1}, ..., x^i _{t - \delta _i}, ..., x^m _{t - \delta _m})^{\rm T}$ is the delayed state vector (note that each component is delayed by a possibly different amount $\delta _i > 0$), and $\bm{\eta}^{\tau}_t = (\eta^{\tau,1} _t, ..., \eta^{\tau,j} _t, ..., \eta^{\tau,n}_t)^{\rm T}$ is a vector of independent noises $\eta^{\tau,j}$, where $\eta^{\tau,j}$ are colored noises (harmonic noises \cite{schimansky1990harmonic}) with characteristic correlation times $\tau_j$, as described in detail in \citet{pesce2013stratonovich}.  We study the limit of this SDE as the parameters $\delta_i$ and $\tau_j$ all go to zero at the same rate, i.e., $\delta_i = c_i \epsilon$, $\tau_j = k_j \epsilon$ where $c_i$ and $k_j$ are constants and $\epsilon \to 0$.  Using a modification of the method outlined in Section~\ref{sec:case:fdr}, we obtain the limiting system
\begin{widetext}
\begin{equation} \label{thm limiting equation 2}
d y^i _t = f^i (\bm{y} _t) dt + 
\underbrace{\sum _{p,j} g^{pj} (\bm{y} _t) \frac{\partial g^{ij} (\bm{y} _t)}{\partial y_p} {1\over2} \left( 1 + \frac{c _p}{k _j} \right)^{-1} dt}_{\mbox{nose-induced drifts}}
+ \sum_j g^{ij} (\bm{y}_t) dW^j_t \; .
\end{equation}
\end{widetext}
The noise-induced drift terms are again of the It\^o-Stratonovich correction type, entering with coefficients that are explicit functions of the $c_i$ and $k_j$. 

\section{Applications, future work and perspectives}\label{sec:outlook}

As we have seen in the previous sections, there is often a need to derive effective and tractable mathematical models that  reduce the number of degrees of freedom of real systems while still representing their complex nature. In fact, the exact modeling of phenomena discussed in this review would require access to their microscopic dynamics, whose time scales are typically much shorter than the observable time scales. A further reduction can be obtained by considering limits in which one or more natural time scales of the problem go to zero. We have also seen that the presence of multiplicative noise (in its multifaceted forms) leads to the appearance of noise-induced drifts in the effective SDEs. Importantly, recent experiments have been able to measure these noise-induced drifts and their consequences in the case of Brownian particles in thermal equilibrium with a heat bath \cite{lanccon2001drift,lanccon2002brownian,volpe2010influence,brettschneider2011force} and in the case of electric circuits \cite{smythe1983observation,pesce2013stratonovich}. Even more importantly, at least one subsequent experiment \cite{mijalkov2015engineering} puts forward a concrete application, by using a noise-induced drift to control the long-term behavior of autonomous agents. 

The explicit formulae for the noise-induced drifts, in particular in the multidimensional case, provide one with powerful tools to study a wealth of interesting phenomena in physics, biology and other fields which use stochastic differential system models. We expect future work to focus on noise-induced drifts and on their dramatic consequences in many cases where SDEs with multiplicative noise are routinely employed to predict the behavior and evolution of complex physical, chemical, biological and economic phenomena. In particular, future work will study in more detail the nature and significance of noise-induced drifts in multidimensional systems. In fact, while several theoretical works have dealt with the multidimensional case, all experiments performed until now focus on noise-induced drifts emerging in effectively one-dimensional systems, i.e., systems where the number of effective degrees of freedom has been reduced to one, even when they are intrinsically multidimensional. For example, in \citet{mijalkov2015engineering} the motion of the robots occurs in a plane, but the effective noise-induced radial drift is measured only along the radial coordinate.

Here we provide a list of topics of interest, focusing on effects that are important for applications and/or that can be verified experimentally.
\begin{itemize}

\item {\bf More realistic experimental model systems.} Electrical circuits are relatively easily controllable physical systems with damping and noise.  As such, they provide a natural class of systems whose parameters can be manipulated to test the theory, e.g., to observe noise-induced bifurcations \cite{smythe1983observation} and transitions \cite{pesce2013stratonovich}. However, they are also relatively simple and uninteresting physical systems. It will therefore be crucial to move towards experimentation in more relevant and realistic systems. Biological systems can be investigated starting from simple bacterial colonies reacting to a time-varying environment in order to study whether, e.g., noise-induced bifurcation in the population dynamics may occur, and moving at a later stage towards more complex ecosystems. Economic systems can be analyzed by using available econometric data; for example, it would be fascinating to study the possibility that booms and bursts in the stock market might be due to a noise-induced transition similar to the one described in Section~\ref{sec:case:color}.

\item {\bf Effect of multiplicative noise on steady-state distributions.} As we have seen in Section~\ref{sec:simple} and, in particular, in Fig.~\ref{fig4}, the presence of a noise-induced drift changes the stationary distribution of an SDE system (if it has one). While for systems satisfying the fluctuation-dissipation relation, e.g., in thermal equilibrium, their potential landscape and steady-state (or, in this case, equilibrium) distribution are connected by the Boltzmann statistics \cite{lau2007state}, this is not necessarily the case for other systems \cite{ao2007existence}. Things become particularly tantalizing when considering multidimensional systems, where also non-conservative forces (e.g. magnetic) may be present. Overall, it will be interesting to explore the interplay between multiplicative noise, noise-induced drifts, non-conservative forces and steady-state probability distributions both from a theoretical and experimental point of view.

\item {\bf Noise-induced bifurcations.} The noise-induced drift can modify the properties of the resulting dynamical system radically, making it undergo a bifurcation. Interestingly, one of the first works studying experimentally systems with multiplicative noise \cite{smythe1983observation} addressed precisely the issue of how the transition from a monostable to a bistable behavior in a noisy electric circuit was affected by the presence of multiplicative noise (Section~\ref{sec:case:color} and Fig.~\ref{fig9}). This is a good starting point to explore more complex situations. In particular, we are planning to identify conditions under which the presence of noise and state-dependent damping induces specific types of bifurcations, e.g., saddle-node or Hopf bifurcations. Furthermore, we are planning to study models of population dynamics, including, e.g., effects of randomness on Lotka-Volterra-type equations. We emphasize that the very interesting theory of \emph{stochastic bifurcations} (see, e.g., \citet[Chapter 9]{arnold2013random}) is concerned with bifurcations of vector fields (or their absence) under adding individual realizations of the noise and thus has a different focus from the one proposed here. Similarly, \citet{arnold1990stabilization} studies stability of equilibria of dynamical systems perturbed by individual noise realizations, rather than modified by the noise-induced drifts considered here.

\item {\bf Noise-induced drifts in thermophoresis.} While we have already suggested that noise-induced drifts might play a role in thermophoresis \cite{hottovy2012thermophoresis}, this is a subject that still needs to be investigated in detail both theoretically and experimentally. In particular, in the presence of a colored noise, the damping term should involve a time delay and one should study the noise-induced drift in the resulting integro-differrential SDE.

\item {\bf Noise-induced drifts in noise-induced phenomena.} Noise plays a crucial (and constructive) role in many phenomena such as Kramers transitions \cite{kramers1940brownian}, stochastic resonance \cite{gammaitoni1998stochastic} and Brownian ratchets \cite{faucheux1995optical}. It will be fascinating to explore how multiplicative noises and noise-induced drifts can affect such phenomena.

\item {\bf Entropy production in the small mass limit.} Entropy production in stochastic systems has been a subject of numerous recent works (a systematic exposition is given in \citet{chetrite2008fluctuation}). \citet{celani2012anomalous} discuss the behavior of entropy production for the equation equivalent to SDE~(\ref{eq:nofdr}) with constant damping (and zero external force), where there is no noise-induced drift.  We propose to study entropy production in the general SDEs~(\ref{eq:SDEgeneral}).  This may lead to a variational characterization of the noise-induced drift.

\item {\bf Noise-induced drifts in curved spaces.} Another direction of future work is concerned with the diffusion of Brownian and active Brownian particles \cite{ebbens2010pursuit,volpe2014simulation} on surfaces. In addition to its intrinsic mathematical interest and beauty, diffusion on surfaces occurs naturally in biology (e.g., molecular complexes on a cell membrane, white blood cells on the surface of an alveolus) and in physics (e.g., colloids trapped on a membrane or interface).  The techniques outlined in this review allow one to study the zero-mass and related limits of equations describing such systems.  In particular, we are planning two theoretical projects.  In the first one, we will present the Wiener process (Brownian motion) on a manifold as a zero-mass limit of an inertial system, justifying its use in mathematical modeling of overdamped systems of surface diffusion.  In the second project, we will consider a particle moving on a two-dimensional surface by inertia, with rapid random changes of direction. Considering an active particle that rotates around its center, we aim to show that in the limit of fast rotations, the particle's dynamics is described by the Wiener process.  To complete these two projects we will couple the techniques presented here with those of stochastic differential geometry \cite{nelson1985quantum,hsu2002stochastic}.  Finally, we are planning a numerical study of diffusion on two-dimensional surfaces in the presence of interesting geometry, resulting in long-term particle trapping, similarly to the results reported (in a different context) by \citet{chepizhko2013diffusion}.

\item {\bf Quantum noise-induced drift.} Another interesting research direction is to study the phenomenon of nise-induced drift in open quantum systems. The dynamics of such systems in the Markov approximation is described by quantum Langevin equations (in the Heisenberg picture). We are planning to conduct an analysis of these equations, similar to the integration by parts technique described here. In some systems, the master equation (analog of the Kolmogorov equation of classical theory) may be more amenable to analysis, patterned in this case on the classical multiscale analysis (homogenization). Among others, we will study quantum Brownian particles, whose coupling to the environment depends on its position. A physical realization of such system is the motion of an impurity atom interacting with a Bose-Einstein condensate.  See also the recent review by \citet{massignan2015quantum} and the references therein.

\end{itemize}
In conclusion, the study of multiplicative noise and of the associated noise-induced drifts is rapidly becoming a fertile field of research. It opens several interesting avenues towards studying new phenomena and offers exciting future research directions.

\appendix

\section{Finite-difference (FD) numerical simulations}\label{app:sim}

The numerical integration of SDEs is discussed in detail in \citet{kloeden1992numerical}; here we provide a primer on how to integrate SDEs with multiplicative noise accounting for the integration convention.

In the FD integration of an ordinary differential equation (ODE), the continuous-time solution $x(t)$ of the ODE is approximated by a discrete-time sequence $x_n$, which is the solution of the corresponding FDE evaluated at regular time steps $t_n = n \Delta t$. If $\Delta t$ is sufficiently short, $x_n \approx x(t_n)$. For example, in the case of a ${\rm 1^{st}}$ order ODE, the FDE is obtained by perfoming the following substitutions:
\begin{eqnarray*}
x(t) & \Rightarrow & x_n \; ,\\
\dot{x}(t) & \Rightarrow & \frac{x_{n+1}-x_n}{\Delta t} \; .
\end{eqnarray*}
The solution is then obtained by solving the resulting FDE recursively for $x_{n+1}$, using the previous value $x_n$ as initial condition.

Let us now consider the SDE
\begin{equation}\label{eq:appendix1}
dx_t = g(x_t) dt + \sigma(x_t) \circ_\alpha dW_t \; ,
\end{equation}
where the noise term is to be integrated with the convention $\alpha$. As we have seen at the end of Section~\ref{sec:simple}, the SDE~(\ref{eq:appendix1}) is equivalent to 
\begin{equation}\label{eq:appendix2}
dx_t = g(x_t) dt + \alpha \sigma(x_t) \frac{d\sigma(x_t)}{dx} dt + \sigma(x_t) dW_t \; ,
\end{equation}
where the multiplicative noise term is an It\^o integral. The numerical integration of the first two terms on the right-hand side of SDE~(\ref{eq:appendix2}) is straightforward and can be performed as for the case of ODEs. In the FDE, the noise term, i.e., $\sigma(x_t) dW_t$, is replaced by $\sigma(x_n) w_n$, where $w_n$ is a Gaussian random number with zero mean and variance $1/\Delta t$. Thus, the resulting FDE corresponding to SDE~(\ref{eq:appendix1}) (and SDE~(\ref{eq:appendix2})) is 
\begin{equation}
x_{n+1} = x_n + g(x_n) \Delta t + \alpha \sigma(x_n) \frac{d\sigma(x_n)}{dx} dt + \sigma(x_n) w_n \; .
\end{equation}
This approach can be straightforwardly generalized to vectorial systems.

\begin{acknowledgments}
We thank our collaborators on various parts of the projects presented here: C. Bechinger, T. Brettschneider, O. Duman, L. Helden, S. Hottovy, A. McDaniel, M. Mijalkov and G. Pesce.
We are also grateful to several colleagues for discussions and important comments: P. Ao, V. Blickle, T. Franosch, A. Gambassi, K. Gaw\c edzki, T. Kurtz, M. Lewenstein, R. Mannella, J. Mehl, M. Rauscher, P. Reimann, U. Seifert, S.R.S. Varadhan and J. Watkins.
GV has been partially financed by Marie Curie Career Integration Grant (MC-CIG) PCIG11 GA-2012-321726 and a Distinguished Young Scientist award of the Turkish Academy of Sciences (T\"UBA).
JW has been partially funded by NSF grant  DMS 131271.
\end{acknowledgments}

\bibliography{biblio}

\end{document}